 \documentclass{aa}
 \usepackage{psfig}

\newcommand{\HP}{{\sc hipparcos}}
\newcommand{\C}{Cepheids}
\newcommand{\M}{{\sc 2mass}}
\newcommand{\OG}{{\sc ogle}}
\newcommand{\DE}{{\sc denis}}

\begin{document}

\thesaurus{03(08.04.1, 08.22.1, 11.13.1, 12.04.3)} 
\title{LMC and SMC Cepheids: combining {\sc ogle} with {\sc
denis} and {\sc 2mass} infrared data
}

\author{M.A.T. Groenewegen 
}

\offprints{Martin Groenewegen (groen@mpa-garching.mpg.de)}

\institute{
Max-Planck Institut f\"ur Astrophysik, Karl-Schwarzschild-Stra{\ss}e 1, 
D-85740 Garching, Germany
}

\date{received: 2000,  accepted: 2000}

\authorrunning{Groenewegen}
\titlerunning{Galactic, LMC and SMC Cepheids: combining {\sc ogle}
with {\sc denis} and {\sc 2mass}}

\maketitle
 
\begin{abstract}

I cross-correlate the \OG\ database of Cepheids in the Large and Small
Magellanic Clouds  (MCs) with the second incremental release of the \M\
$JHK_{\rm s}$ survey, and the {\sc denis} $IJK_{\rm s}$ point source
catalog towards the MCs.

Of the 3384 \C\ in \OG\ in both Clouds, 1745 have a counterpart in the
\M\ survey within an 1\arcsec\ radius and good-photometry in all three
bands. Only 173 have a counterpart in the \DE\ survey within an
1\arcsec\ radius and good-photometry in all three bands. The reason
for this difference is that the limiting magnitudes of the \M\
survey are considerably fainter than for the \DE\ survey.

The standard stars of the Carter IR photometric system are also
correlated with the \M\ survey to derive transformation equations from
the natural \M\ system to the Carter-system.

In order to describe to first order the geometrical configuration of
the MCs a plane is fitted to the reddening-free Wesenheit index ($W =
I - 1.55 \times (V-I)$) and the inclination and position angle of the
line-of-nodes are determined. For the LMC an inclination angle of
18 $\pm$ 3\degr\ is derived, somewhat smaller that previous determinations.
For the SMC a value of $i = 68 \pm 2$\degr\ is derived in good
agreement with previous determinations. These results are being used
to take out the effect that some \C\ are closer to the observer than
others. For the LMC this effect is small over the area covered by the
\OG\ survey, but for the SMC $PL$-relations with a reduced scatter are
derived in this way. Nevertheless, the dispersion in the $PL$-relation
in $W$ for the SMC is larger than that for the LMC, indicating a
significant intrinsic depth of the SMC of about 14 kpc
(front-to-back). The depth of the LMC is found to be small compared
to the intrinsic scatter in the $PL$-relation.

Using single-epoch \M\ data for the \OG\ \C\, $PL$-relations are fitted in
$JHK$ (on the Carter system) for both Fundamental mode (FU) and
First-overtone (FO) pulsators, taking out the effect of the inclined
disk.  Because of the large number of stars available, this results in
the most accurately determined slopes in the infrared up to now, and
slopes for FO pulsators derived for the first time. This provides
additional constraints for theoretical models.

The 16 derived $PL$-relations (in $WJHK$ for FU and FO pulsators for
LMC and SMC) are used to derive a relative distance modulus
${\Delta}_{\rm SMC-LMC} = 0.50 \pm 0.02$ assuming no metallicity
correction. Two recent theoretical models predict different
metallicity corrections, indicating a systematic uncertainty of up to
0.1 mag.

Using the database of Galactic \C\ observed with \HP\ and the method
outlined in Groenewegen \& Oudmaijer (2000), the zero points of
Galactic $PL$-relations for FU pulsators in $W$ and $K$ are derived,
using the newly derived slopes for LMC and SMC pulsators.

Combining the zero points of the Galactic and MCs $PL$-relations
distance moduli of 18.60 $\pm$ 0.11 (based on $W$) and 18.55 $\pm$
0.17 (based on $K$) to the LMC, and of 19.11 $\pm$ 0.11 ($W$) and 19.04
$\pm$ 0.17 ($K$) to the SMC are derived (without taking into account 
possible metallicity corrections).

\keywords{Stars: distances - Cepheids - Magellanic Clouds - distance scale}

\end{abstract}

\section{Introduction}

Cepheids are important standard candles in determining the
extra-galactic distance scale. The results of the \HP\ mission allowed
a calibration of the Galactic Period-Luminosity ($PL-$) relation in
the $V$-band (Feast \& Catchpole 1997), $V$ and $I$-bands (Lanoix et
al. 1999a) and $V$, $I$, $K$ and the reddening-free Wesenheit index
(Groenewegen \& Oudmaijer 2000, hereafter GO00). Feast \& Catchpole
(1997) and GO00 combined these data with available data on LMC
Cepheids to determine the distance to the LMC.  It is well know that
the infrared has some considerable advantages over the optical region
for deriving distances. First of all the problem of reddening is less
important, and second, as summarised in GO00, the effect of
metallicity on the $PL$-relation seems to be less.  Of the 236 \C\ in
the sample considered in GO00, only 63 have sufficiently well
determined infrared light curves to determine accurate intensity-mean
magnitudes (also see Groenewegen 1999). In addition, the number of
Magellanic Clouds (MCs) \C\ with well determined infrared lightcurves
is also rather small, of order 20 in each Cloud (Laney \& Stobie,
1994, 1986a).

The micro lensing surveys have increased the number of known \C\ in
the MCs tremendously and they provide optical photometry for many
of them. In particularly, Udalski et al. (1999b,c; hereafter
U99b,c\footnote{In April 2000 the \OG\ team released a re-reduced
dataset changing the photometry slightly, and hence the
$PL$-relations with respect to those in the papers published in 1999. 
The new data and corresponding $PL$-relations can be
found on the \OG\ Homepage. When appropriate I will refer to these new
values.})  recently published the results on Cepheids in the Small and
Large Magellanic Cloud from the \OG\ survey.  Complementary, in the
infrared, the \M\ survey is an ongoing all-sky survey in the
$JHK_{\rm s}$ near-infrared bands, that, when completed, will contain
of order 300 million stars (Beichman et al. 1998). The \M\ team
released their second incremental data set in March 2000, that
includes the MC area. In parallel, the {\sc denis}
survey is a survey of the southern hemisphere in $IJK_{\rm s}$
(Epchtein et al. 1999), that recently released a Point Source Catalog
(PSC) of sources in the direction of the MCs (Cioni et al. 2000a).

In this paper the \OG\ database is correlated with the \M\ second
incremental database and the \DE\ PSC of MC stars
(Sect.~2), transformation formulae from the natural \M\ system to the
Carter (1990) system are derived (Sect.~3), the spatial structure of
the MCs is investigated (Sect.~4), and $PL$-relations in $JHK$ for
fundamental mode and first overtone pulsators derived (Sect.~5). The
results are discussed in Sect.~6.

\section{\OG, \M\ and \DE\ Sample selection}

U99b,c describe the dataset regarding \C\ in the direction of the LMC
and SMC, obtained in course of the {\sc ogle-ii} micro lensing
survey. Twenty-one fields in the central parts of the LMC, and 11
fields in the central parts of the SMC of size
14.2\arcmin$\times$57\arcmin\ each were observed in $BVI$, with an
absolute photometric accuracy of 0.01-0.02 mag. U99b,c present tables
per field, with (amongst other items) identification number, pulsation
period, $BVI$ photometry, $E(B-V)$ and pulsation mode. The $E(B-V)$
value was derived by them using red clump stars, and the pulsation
mode was derived by them from two Fourier parameters derived from the
$I$-band lightcurve (see U99b,c for details).  These data were
obtained from the \OG\ homepage
(http://sirius.astrouw.edu.pl/$^{\sim}$ogle/) in April 2000.

As noted in U99b,c there are about a 100 objects in each cloud that
appear in 2 or more fields. They provide a list with the relevant
field and identification number, but I did this positional correlation
independently. Values for the photometry, period and $E(B-V)$ were
averaged where appropriate.  The final number of Cepheids detected by
\OG\ in the LMC is 1335, and in the SMC 2048.

The \M\ survey is an ongoing single-epoch all-sky survey in the
$JHK_{\rm s}$ near-infrared bands. On March 2, 2000 the \M\ team
released the second incremental data set, that includes the MC area.  The easiest way to check if a star is included in this
release is by uplinking a source table with coordinates to the \M\
homepage.  Such a table was prepared including the 1335 LMC and 2048
SMC Cepheids in \OG\ and 236 \C\ in the \HP\ catalog considered in
GO00.

The correlation between the source list and the \M\ database was at
first instance done using a search radius of 5\arcsec.
Inspection of the match between the \OG\ and \HP\ \C\ and \M\
indicated a much better positional coincidence. This is to be expected
as the \M\ positional accuracy is $<$0.2\arcsec\ (Cutri et al. 2000)
and the internal accuracy of the \OG\ database is about 0.15\arcsec\
with possible systematic errors of $<$0.7 \arcsec\ (U99b,c). In quite
a few cases there was a second \M\ source (not the counterpart) within
the search radius of 5\arcsec. The match between \M\ and \HP\ was
excellent with the largest positional difference being 0.56\arcsec. In
view of this, a second and final correlation was made for the \HP\ and \OG\ \C\
using a search radius of 1\arcsec.  Seventy-two correlations were
found with the \HP\ \C, 1511 with the \OG\ SMC, and 894 with the \OG\
LMC \C. The small number of correlations with \HP\ is due to the fact
that the \M\ survey is not yet complete in sky coverage.

On this \M\ dataset a further selection was performed to retain only
those sources with reliable $JHK$ photometry in all three bands. It is
worthwhile to recall here that the ``rd-flg'' in the \M\ database
gives essential information on the quality of the magnitudes (Cutri et
al. 2000). In particular, objects where the rd-flg is 0 (indicating an
upper limit) or 3 or 8 (indicating saturation) in any of the three
bands are eliminated. The final \M\ database of \C\ consists of 52
\HP\ \C\ (not discussed further),
825 \OG\ LMC and 920 \OG\ SMC \C. The \M\ and \OG\ coverage of the LMC
are not identical, and no \M\ sources are found in \OG\ fields
LMC-SC6, 7 and 21.

The \DE\ survey is a survey of the southern sky in $IJK_{\rm s}$
(Epchtein et al. 1999). In April 2000, a point source catalog was
released containing sources and single-epoch photometry in the
direction of the MCs (Cioni et al. 2000a). A subset of 264347 LMC and
69829 SMC sources detected in all three band was obtained using {\sc
ftp} from www.strw.leidenuniv.nl (cd /pub/ldac/dcmc). These were
correlated against the \OG\ survey on position. As both \OG\ and \DE\
contain $I$-band data it was possible to verify how the average
difference between the \OG\ and \DE\ $I$-data depends on the search
radius. Since the \OG\ data contain intensity-mean magnitudes, but the
\DE\ data are single-epoch data, one has to allow for variability. For
the SMC, and a search radius of 2\arcsec, 93 cross-correlations were
found. The average difference in position is 0.68\arcsec\, and the
mean absolute difference in $I$ is 0.18 mag. However, for the 21 stars
with a positional difference larger than 1\arcsec, 9 (or 43\%) have a
difference in $I$ of larger than 0.8 mag, which is a typical value for
the amplitude for a Cepheid in the $I$-band. For the stars with a
positional differences less than 1\arcsec, this is only 5 out of 67 or
7\%.  Given the internal accuracy of the positions in the \DE\ survey
of 0.5\arcsec\ (Epchtein et al. 1999), a search radius of 1\arcsec\
seems appropriate to reduce the number of spurious associations, which
is supported by the large fraction of highly deviant $I$-band data
when a larger search radius is used. As discussed below some true
identifications are missed in this way however. Also some false
coincidences are undoubtedly included (as for the correlation of \OG\
with \M) but these will be removed using sigma-clipping when deriving
$PL-$relations. Using a search radius of 1\arcsec\ the number of
correlations between \DE\ and \OG\ is 71 in the SMC, and 102 in the
LMC, for stars with good photometry in all three bands

It was verified whether all correlations between \DE\ and \OG\ were also
found in the match between \M\ and \OG. Of the 71 SMC \OG\ + \DE\
stars, 61 are also found in \M. In three of the 10 cases where there
is no match this was due to the positional difference which was
$>1$\arcsec\ between \OG\ and \M.  In the other 7 cases it is less
clear, possibly related to the variable nature of the objects. Of the
102 LMC \OG\ + \DE\ stars, 80 were also found in \M. In one of the 22
other cases this was due to a positional difference of $>1$\arcsec\
between \OG\ and \M, and $<1$\arcsec\ between \OG\ and \DE.  One case
is puzzling, and the other 20 cases are all in \OG\ fields LMC-SC6, 7,
21, which apparently have not been covered by \M.

The number of correlations beween \DE\ and \OG\ is much less than the
number of correlations between \OG\ and \M. Some tests using \DE\ data
of sources detected in $I$ and $J$ only, showed that this is mainly
due to the lack of detections in $K$. Some tests in a few \OG\ fields
showed that there may be an equal number of \C\ in \OG\ with \DE\ $I$
and $J$ photometry. In any case it is clear however, that for the
purpose of determining $PL$-relations in the infrared using the
largest possible samples, the \M\ survey is of more interest, because
it reaches fainter magnitudes.

In Appendix A a comparison is presented between the \DE\ $I$ and \OG\
$I$ and \DE\ $JK$ and \M\ $JK$ photometry. This has its limitations
since the comparison is done using variable stars. Yet it may still be
of some interest to other workers, as such a comparison has not been
done yet.

\begin{figure*}
\centerline{\psfig{figure=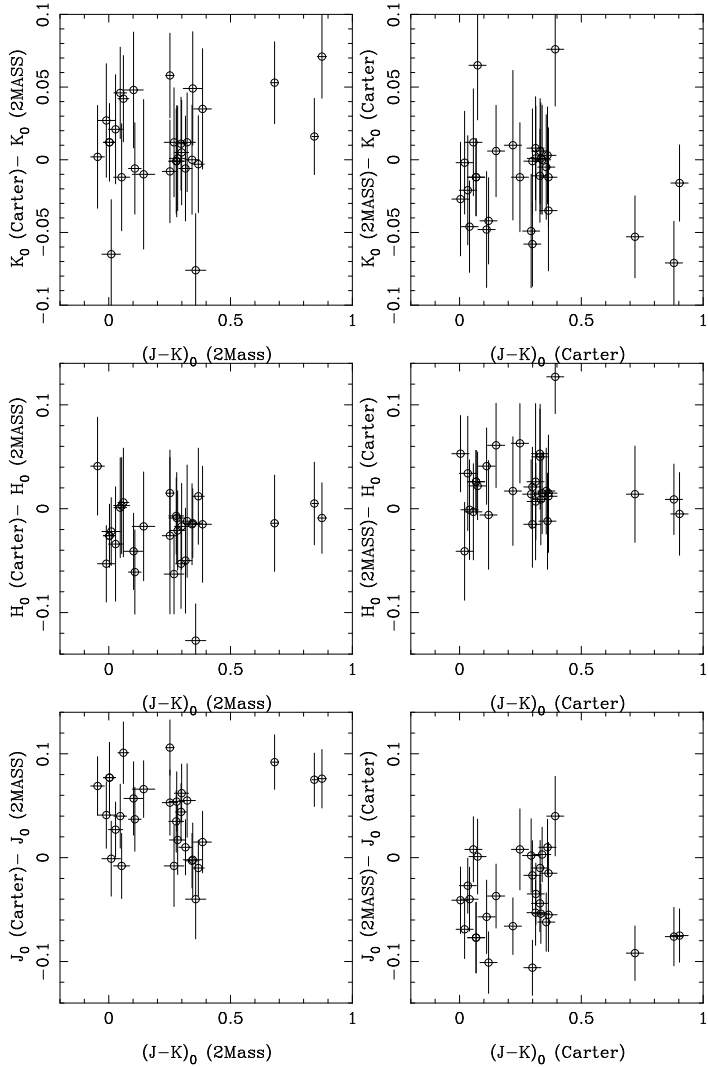,width=9.5cm}}
\caption[]{Comparison between $JHK$ in the Carter (1990) and the
natural system of \M.
}
\end{figure*}

\section{Transformation of the photometric system}

The photometry presented by the \M-team is on the natural system.
However, many of the extisting infrared photometry on both Galactic
(Laney \& Stobie 1992, Groenewegen 1999) and MC Cepheids
(Laney \& Stobie 1994) is on the Carter (1990) system or has been
transformed to that system. It is therefere imperative to estimate any
systematic differences between the Carter and the natural \M\ system.

From Carter (1990) and Carter \& Meadows (1995) 116 stars with $K > 5$
were taken, and a table with coordinates was uplinked to the \M\
homepage to perform a search on position using a search radius of
2\arcsec. Thirty-four matches were found; the largest positional
difference was 1.2\arcsec. However, not all entries are useful as
some of the brighter stars still saturate the \M\ detectors in one or
more bands, despite the a-priori exclusion of stars with $K < 5$. 
The vast majority of stars in \M\ have ``default magnitudes
obtained using profile-fit photometry performed simultaneously on the
combination of all six individual 1.3-seconds ``Read 2-Read 1''
(R2-R1) exposures'' (Cutri et al. 2000).  These sources have rd-flg =
2 in the appropriate band.  Sources brighter than 7-8 magnitudes will
saturate in the 1.3-s R2-R1 exposures. These objects have default
magnitudes from aperture photometry performed on the 51-millisec R1
frames. Such sources have rd-flg = 1 in the appropriate band. Stars
brighter than approximately fifth magnitude will saturate in even the
51-millisec exposures. The rd-flg value for the appropriate band is 3.

Among the 34 sources, there are 15 sources with rd-flg = 111, 10 with
rd-flg = 112, 1 with rd-flg = 122, 4 with rd-flg = 222 and 4 sources
with rd-flg = 3 in atleast one of the bands. The latter were not
considered.

Figure~1 plots the differences ``Carter - 2Mass'' versus $(J-K)$
(2Mass), and ``2Mass - Carter'' versus $(J-K)$ (Carter) in $JHK$. For
these nearby stars reddening is neglected. The errors in the
individual magnitudes are taken from the \M\ database (typically
between 0.01 and 0.03 mag), or are 0.025
mag in the case of the Carter system following the estimate in Carter
\& Meadows (1995). Linear least-square fits show that the significance
of the derived slopes is less than 1$\sigma$ in $HK$ and about
1.5$\sigma$ in $K$, indicating that there is at present no indication
of a colour term in the transformation from the \M\ to the Carter
system.  The average off-sets are $K$(Carter) $-$ $K$(2Mass) = +0.017 $\pm$
0.008, $H$(Carter) $-$ $H$(2Mass) = $-0.021$ $\pm$ 0.009 and
$J$(Carter) $-$ $J$(2Mass) = +0.048 $\pm$ 0.007 (internal errors).  The
dispersion in all three plots is about 0.03 mag. This analysis indicates (A)
that there might be small off-sets between the two infrared systems,
and (B) that the absolute photometric calibration of both is uncertain
at the 0.01-0.02 mag level.  This issue may addressed in more detail when
the \M\ survey has been completed and this analysis may be repeated
using a larger number of stars.

Regarding the same issue, Nikolaev \& Weinberg (2000) note that the \M\
system is similar to the CIT/CTIO system (Elias et al. 1982, Persson
et al. 1998) except for the $K_{\rm s}$ band. They note that the
difference of the absolute value of $(K_{\rm CIT} - K_{\rm s})$ is
less than 0.05 mag.

\section{The spatial structure of the LMC and SMC}

In this section the spatial structure of the MCs is being investigated
in terms of the usual approximation of a flat disk, using the largest
possible dataset, that is the \OG\ sample (without the correlation
with \M). The secondary aim is to correct for any depth effects in order to
obtain tighter $PL$-relations.

Cepheids have previously been used to determine the spatial structure
of the MCs, for example Caldwell \& Coulson (1986) whose model for the
LMC is a plane of insignificant thickness inclined by 29 $\pm$ 6\degr\
with the closest part at position angle (P.A.) 52 $\pm$ 8\degr.
For the SMC they find that a planar model is barely sufficient to
describe this galaxy, and derive an inclination of 70 $\pm$ 3\degr\
with the closest part at P.A. 58 $\pm$ 10\degr. Laney \& Stobie
(1986a) reach similar conclusions and derive $i = 45 \pm 7$ for the LMC
with the nearest part at P.A. 55 $\pm$ 17\degr.  Values for
inclination and position angle derived from other methods and a more
general discussion can be found in the monograph by Westerlund (1997).
Most recently, Weinberg \& Nikolaev (2000) used star counts from the
\M\ survey of the LMC, and a model for the number density of stars, to
derive inclinations between 24 and 28 degrees, and position angles of
the line-of-nodes between 169 and 173 degrees depending on the
population, and an inclination of 42.3 $\pm$ 7.2\degr\ from
carbon-rich Long Period Variables (LPVs).

The coordinate transformations that are necessary have conveniently
been written down in Appendix A.1 in Weinberg \& Nikolaev (2000). To
summarise, for every Cepheid under consideration one knows its right
ascension and declination, and one can derive its distance $r$ (see
below). This coordinate system ($\alpha$, $\delta$, $r$) is
transformed into a rectangular coordinate system ($x_0$, $y_0$,
$z_0$), which has its origin at the center ($\alpha_0$, $\delta_0$,
$R$) of the LMC or SMC, the $z_0$-axis towards the observer, the
$x_0$-axis anti-parallel to the right ascension axis, and $y_0$
parallel to the declination axis. Another rectangular coordinate
system is introduced ($x^\prime$, $y^\prime$, $z^\prime$), which is
rotated about the $z_0$-axis by an angle $\theta$ counterclockwise
(counting East of North), and about the new $x^\prime$-axis (the
line-of-nodes) by the inclination angle $i$ clockwise (that is
rotation from the new $y$-axis towards the $-z_0$-axis). Note that
there seem to exist slightly different definitions in the literature
regarding this issue, related to, for example, the definition of
negative/positive inclination\footnote{In particular, using the
present notation, the result of Caldwell \& Coulson (1986) implies
$\theta = 232 \pm 8$, 
the result of Laney \& Stobie (1986a) implies $\theta = 235 \pm 17$,
and that of Weinberg \& Nikolaev (2000) implies a P.A. of the
line-of-nodes between 259 and
263 degrees for the LMC.}.

The above procedure assumes one knows the distance to the center of
the galaxy, and the distance to the individual Cepheids. To this end
an observed $PL$-relation is used, together with a typical distance
(the assumed value for $R$ is arbitrary because one basically look at
residuals in magnitude space relative to the observed mean
$PL$-relation. However, for convenience, it is preferred here to work
in physical, rather than magnitude, space, and therefore an arbitrary
mean distance to the center of the galaxy has to be adopted).

In particular, the $PL$-relation using the reddening-free
Wesenheit-index using the $V$ and $I$ colours is being used as the
reference $PL$-relation. This is done as the observed scatter in this
relation is less than in the $PL$-relations in $V$ or $I$, and because
the Wesenheit-index, by construction, does not depend on the assumed
value for $E(B-V)$. It depends only on the reddening law. Its
definition is $W$ = $I - 1.55 \times (V-I)$ for the \OG\ filter system
and the extinction law by Cardelli et al. (1989) (Udalski et
al. 1999a; hereafter U99a)

\subsection{The LMC}

U99a derived the zero point and slope of the $PL$-relation in $W$, for
periods longer than about 2.5 days and fundamental pulsators (FU), and
applying clipping at the 2.5$\sigma$ level (see Table~1, entry 1).
They used this cut-off in period as in the SMC there is a change of
slope at shorter periods (Bauer et al. 1999). Using the same criteria
I derive almost identical results (solution 2), altough I would prefer
to use a less stringent clipping to have a larger sample.  A solution
with clipping at the $4\sigma$ level is also included (solution 3).

The change of slope observed in the {\sc eros} data for the SMC was
not observed in the LMC (Bauer et al. 1999), and also theoretically,
one would expect this change of slope to occur at shorter periods for
higher metallicity (Alibert et al. 1999).  Table~1 also includes
a fit using no constraint on the period (solution 4). The results are
nearly identical, but obtained for a larger sample.

U99a only briefly mention the result for the slope derived for First
Overtone (FO) pulsators, and, in fact, more details are given on the
\OG\ Homepage (entry 5 in Table~1), which has
been derived for stars with $\log P > -0.2$ (Udalsky, private
communication). I find essentially the same result (solution 6), also
using a less stringent 4$\sigma$ clipping (solution 7). However, for
consistency, a cut-off at $\log P = 0.4$ for FU pulsators should
correspond to a cut-off at about $\log P = 0.25$ for FO pulsators,
according the usual ratio of $P_0/P_1$ (for example Feast \& Catchpole
1997). Table~1 gives the solutions for stars above and below this
cut-off in period (solutions 8, 9). The slopes are nearly identical,
but there is a significant off-set in zero point.  Bauer et al. (1999)
for a smaller sample of 113 FO LMC Cepheids detected by {\sc eros}
found that both zero point and slope did not depend on a cut-off in
period. However their error bars are roughly 4 times larger than here,
and so their sample may not have been large enough to detect this effect.

\begin{table*}
\caption[]{$PL$-relations in the LMC of the form $M = a \times \log P +b$}
\begin{tabular}{ccccccl} \hline
solution & $M$  &   $a$     &     $b$     & $\sigma$ & N  & Remarks \\ \hline

1& $W$  &  $-3.300 \pm 0.011$ & 15.868 $\pm$ 0.008 & 0.058 & 668 & 
                        U99a; FU; $\log P>0.4$; 2.5$\sigma$ clipping \\

2& $W$  &  $-3.303 \pm 0.012$ & 15.869 $\pm$ 0.008 & 0.059 & 639 & 
                  This paper; FU; $\log P>0.4$; 2.5$\sigma$ clipping \\

3& $W$  &  $-3.310 \pm 0.014$ & 15.871 $\pm$ 0.009 & 0.073 & 673 & 
                  This paper; FU; $\log P>0.4$; 4$\sigma$ clipping \\

4& $W$  &  $-3.320 \pm 0.013$ & 15.880 $\pm$ 0.009 & 0.074 & 713 & 
                  This paper; FU; all $\log P$; 4$\sigma$ clipping \\

5& $W$  &  $-3.425 \pm 0.017$ & 15.380 $\pm$ 0.006 & 0.060 & 457 & 
                        U99a; FO; $\log P>-0.2 $; 2.5$\sigma$ clipping \\

6& $W$  &  $-3.434 \pm 0.017$ & 15.383 $\pm$ 0.006 & 0.060 & 424 & 
                  This paper; FO; $\log P>-0.2 $; 2.5$\sigma$ clipping \\

7& $W$  &  $-3.400 \pm 0.018$ & 15.369 $\pm$ 0.007 & 0.072 & 450 & 
                  This paper; FO; all $\log P$; 4$\sigma$ clipping \\

8& $W$  &  $-3.300 \pm 0.032$ & 15.323 $\pm$ 0.013 & 0.068 & 344 & 
                  This paper; FO; $\log P>0.25$; 4$\sigma$ clipping \\

9& $W$  &  $-3.266 \pm 0.050$ & 15.385 $\pm$ 0.009 & 0.084 & 108 & 
                  This paper; FO; $\log P<0.25$; 4$\sigma$ clipping \\

10& $W$  &  $-3.337 \pm 0.013$ & 15.890 $\pm$ 0.008 & 0.072 & 713 & 
         FU; all $\log P$; 4$\sigma$ clipping; planar effect taken out \\

11& $W$  &  $-3.398 \pm 0.018$ & 15.370 $\pm$ 0.007 & 0.072 & 450 & 
         FO; all $\log P$; 4$\sigma$ clipping; planar effect taken out \\

\hline
\end{tabular}
\end{table*}

For the investigation of the effect of an inclined disk the FU 
pulsators are being considered first, using the relevant observed
$PL$-relation (solution 4). An arbitrary distance to the center of the
LMC of $R = 50$ kpc is assumed, and the center of the disk 
is assumed to be given
by the mean of the 713 \C\ in the sample: $\alpha_0 = 81.384$,
$\delta_0 = -69.782$ degrees (J2000). The observed $PL$-relation is
then transformed into an absolute relation $M_{\rm W}$ = $-3.320$  $\log
P$ $-2.615$, and with this the distance to every Cepheid can be calculated,
and the coordinates ($x_0$, $y_0$, $z_0$) and ($x^\prime$, $y^\prime$,
$z^\prime$) determined for an assumed position angle and inclination.

There may be one caveat of working in real space rather than
magnitude space that should be pointed out here. The dispersion around
the observed $PL$-relation is due to three effects. The depth effect
along the line-of-sight, or, equivalently, the distance $z_0$ to the
plane of the sky, photometric errors, and ``cosmic'' scatter. The
latter term arises because a linear $PL$-relation is an approximation,
metallicity effects, or other effects. The photometric errors are small
(of order 0.01 mag) and the other two effects are difficult to
distinguish, and when considering correlations of the residual
magnitudes with respect to the mean $PL$-relation with position need
not to be distinguished.  When working in physical space, the $z$
values derived are {\bf NOT} the actual distances below or above the
mid-plane. In fact the values for the distance to the mid-plane are
upper limits as the ``cosmic'' dispersion is non-zero.

Figure~2 shows the distributions for $i = \theta = 0$. The upper-left
panel shows in filled squares stars that are closer to the observer
(that is brighter than the $PL$-relation) and in open circles \C\ that
are further away (fainter than the $PL$-relation)\footnote{In the
electronic version, this is highlighted by using different colour
schemes as well.}.  The other 3 panels show the projection on to
different planes. From the top right panel a rotation angle of
approximately 26 (c.q. 206) or 116 (c.q. 296) degrees is derived. The
question of the inclination is not immediately clear. Figure~3 shows
the distribution and projections for $\theta = 26$ and $i = 0$. The
$x^\prime$-axis now runs along the major axis from bottom left to top
right and $y^\prime$-axis runs along the minor axis from bottom right
to top left. The $x^\prime$ and $y^\prime$ coordinates are now
uncorrelated, by construction. Their is no correlation between
$z^\prime$ and $y^\prime$, indicating no rotation over the
$x^\prime$-axis. The correlation between $z^\prime$ and $x^\prime$
indicates a rotation along the $y^\prime$-axis by 18 degrees with the
south-east part tilted towards us. Given the definition of the
coordinate system, and considering the errors in the slopes, this
implies a position angle of the line-of-nodes of $\theta = 296 \pm
0.5$, $i = 18 \pm 3$\degr.

\begin{figure}
\centerline{\psfig{figure=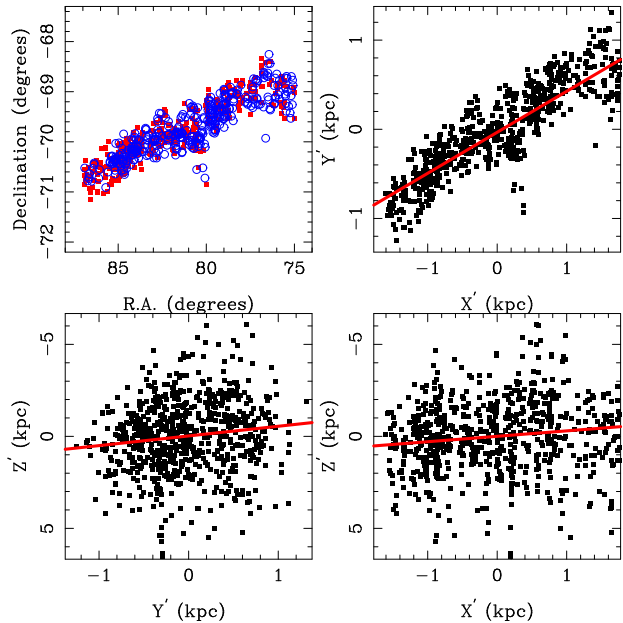,width=8.5cm}}
\caption[]{Distribution on the sky of FU \C\ who are brighter (filled
squares) or fainter (open circles) than the mean $PL$-relation (upper
left panel), and projection in ($x^\prime, y^\prime, z^\prime$)
coordinates for $i = 0$ and $\theta = 0$. Solid lines indicate the
results of linear least-square fits.
}
\end{figure}

\begin{figure}
\centerline{\psfig{figure=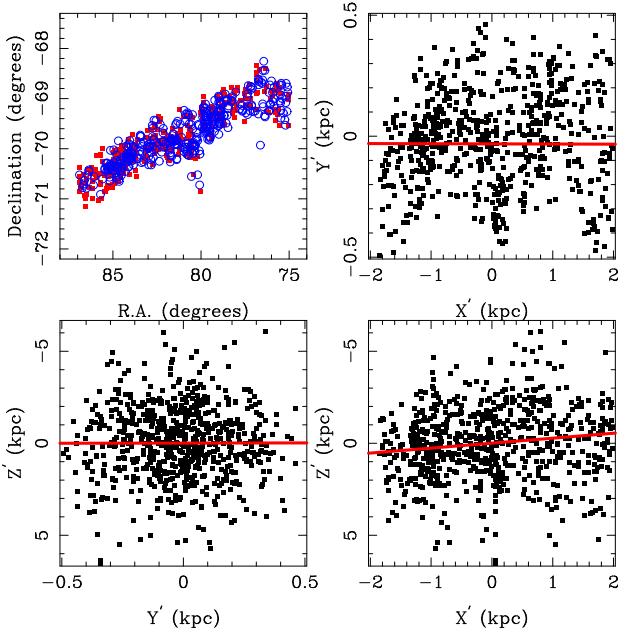,width=8.5cm}}
\caption[]{As Fig.~2 for $i = 0$ and $\theta = 26\degr$.
}
\end{figure}


This analysis was repeated using all FO pulsators (solution 7), and
Fig.~4 shows the result for  $i = 0$ and $\theta = 24$. The
inclination implied by the slope in the  $z^\prime$ and $y^\prime$
plane is not significant being 16 $\pm$ 17\degr, and the slope in the
other plane of 10 $\pm$ 5\degr\ is consistent with that derived from the
FU pulsators.

\begin{figure}
\centerline{\psfig{figure=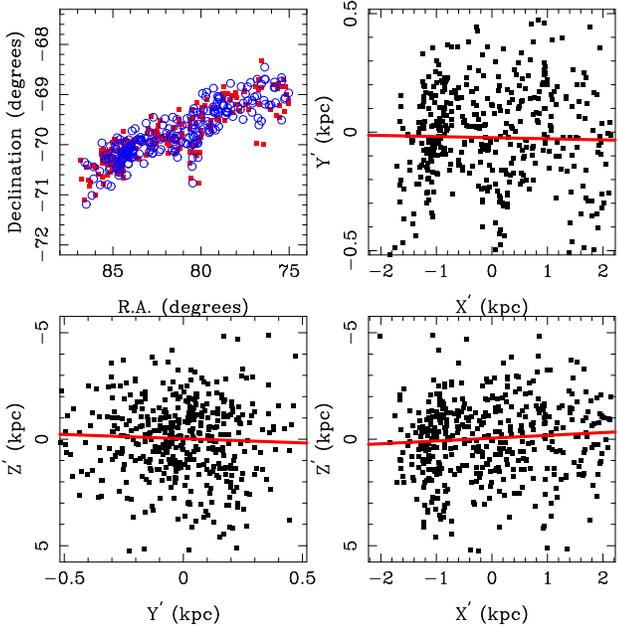,width=8.5cm}}
\caption[]{As Fig.~2 for $i = 0$ and $\theta = 24\degr$ and FO pulsators.
}
\end{figure}

These results are in fair agreement with previous results in the
literature, where larger inclinations are usually quoted (see the
beginning of Sect.~4.0). Probably the main reason is that those
studies covered a much larger area, and not only the bar region. In
particular it would be extremely useful to have \OG\ fields at some
distance from the bar, along P.A. of about 30, 120, 210 and 300
degrees. One other possibility is that the structure of the LMC is
truly different in the central region from the more global structure.
A very recent preprint (Zhao \& Evans 2000) eluded to the possibility
of an off-setted and misaligned LMC bar with respect to the LMC disk.

As a further illustration Fig.~5 depicts the distance from the
mid-plane assuming no intrinsic scatter of the $PL$-relation. In fact,
this plot illustrates the upper limit to this distribution, and the
true distribution ($x$-scale) should be compressed by a factor of
about 9 (see below).

From McCall (1993) it is estimated that the layer containing 99\% of
young stars has a thickness of 1.0 $\pm$ 0.3 kpc, for a distance to SN
1987A of 51.2 $\pm$ 1.2 kpc (Panagia 2000). The half-thickness of the
layer containing 50\% of the young stars can then be calculated to be
0.1 $\pm$ 0.03 kpc. McCall also derives $H/R_0 = 0.120 \pm 0.031$,
where $H$ is the vertical scale height and $R_0$ the radial scale
length of the LMC disk. For $R_0$ between 1.4 and 1.8 kpc derived from
\M\ data (Weinberg \& Nikolaev 2000), this results in $H = 0.19 \pm
0.06$ kpc.  Taking this latter value, it is concluded that at the
distance of the LMC the thickness of the LMC disk gives rise to a
1$\sigma$ dispersion in $PL$-relations of approximately 0.008 mag
only. The total front-to-back depth corresponds only to 0.043 mag.

A final remark regarding this topic is that the theoretically
predicted dispersion in the Wesenheit-index is 0.10 mag (Bono \&
Marconi, 2000, private communication). This is in fact slightly larger
than the observed one. However, this is again consistent with the
notion that the intrinsic depth of the LMC plays no role in the
observed scatter in the $PL$-relations towards the LMC.

Weinberg \& Nikolaev (2000) recently claimed the LMC to be
geometrically thick along the line of sight with a thickness of
$\sim$14 kpc. They used stars in a color selected region based on \M\
data and assumed these to be carbon-rich LPVs. They derive widths of
the magnitude distribution functions of 2$\sigma$ = 0.7 mag, while
they claim their standard candles to have an intrinsic spread of 0.2
mag, from which they derive a front-to-back thickness of $\sim$14 kpc,
after they rejected other possible explanations. For a similar
population, Bessell et al. (1986), from the intrinsic
line-of-sight velocity dispersion of old LPVs, derived a vertical scale
height $H \sim$0.3 kpc, consistent with the notion of a thin disk. \\


\begin{figure}
\centerline{\psfig{figure=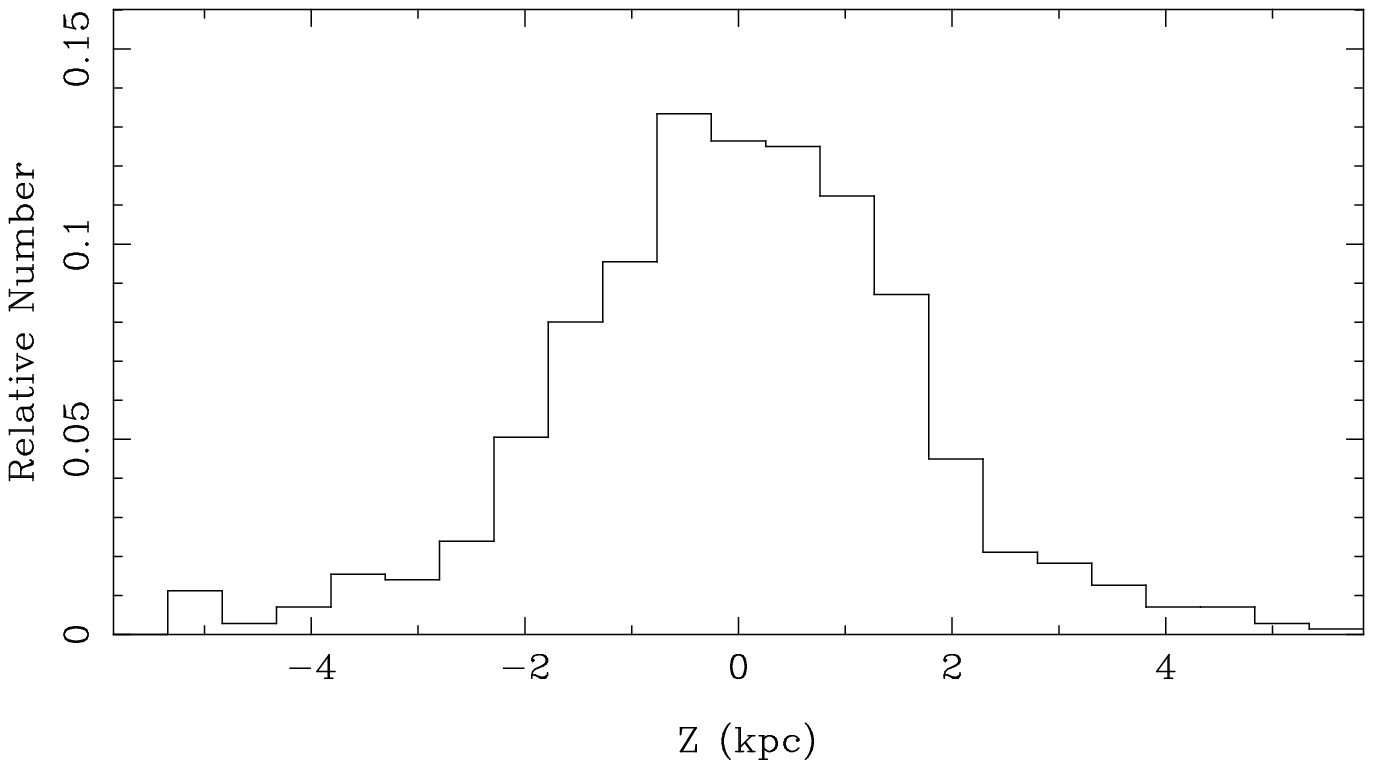,width=8.5cm}}
\caption[]{If the intrinsic spread of the $PL$-relation were zero,
this would be the distribution of stars along the line-of-sight towards the
LMC.  Positive-$z$ are towards the observer. As the intrinsic spread
is in fact 0.008 mag, the horizontal scale should be compressed
in such a way that the 4 kpc thick actually correponds to 450 pc. }
\end{figure}

Although it will turn out academic for the LMC, it will be discussed
now, how an improved $PL$-relation can be derived once position angle
and inclination have been determined. If the approximation of a thin
disk is correct, one can calculate those distances, $r^\prime$,
respectively, $\tilde{r}$, that make $z_0 = 0$ and $z^\prime = 0$ (for
fixed $\alpha$ and $\delta$ and given $\theta$ and $i$). That is, for
every star one can subtract the correction in magnitude, 5 $\log(
\tilde{r} / r^\prime )$, to put all stars in the plane of the sky, and
then fit the $PL$-relation again. Planar effects are in this way taken
out.  The result is listed in Table~1 as solutions 9 and 10, and are
essentially the same as solutions 4 and 7 with only marginally lower
errors and the fit coefficients.  This indicates that over the region
covered by \OG, depth effects are not important, and that the width of
the $PL$-relation is intrinsic. The resulting final $PL$-relations are
shown in Figs.~6 and 7.

\begin{figure}
\centerline{\psfig{figure=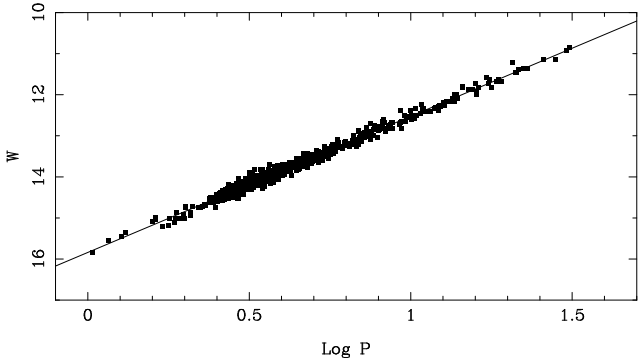,width=8.5cm}}
\caption[]{Final $PL$-relation in $W$ for FU mode pulsators in the LMC
when the depth effect has been taken out. The drawn line is the best
fit.  }
\end{figure}

\begin{figure}
\centerline{\psfig{figure=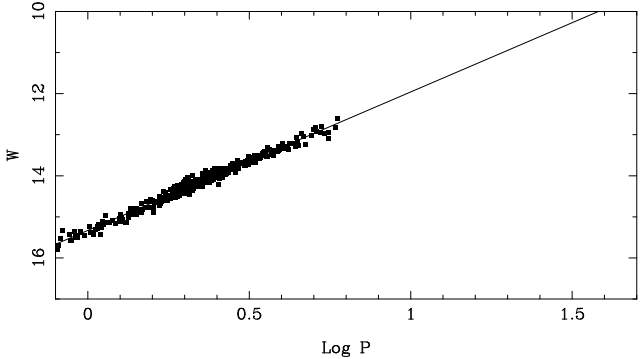,width=8.5cm}}
\caption[]{As Fig.~6 for FO mode pulsators in the LMC.}
\end{figure}

\subsection{SMC}

This section is largely a repeat of the previous subsection but for
the SMC. U99a derived the zero point and slope of the $PL$-relation in
$W$, for periods longer than about 2.5 days and fundamental pulsators,
and applying clipping at the 2.5$\sigma$ level (see Table~2, solution
1).  They used this cut-off in period as there is a change of slope
observed at shorter periods (Bauer et al. 1999). Using the same
criteria I derive almost identical results (solution 2). A solution
with clipping at the $4\sigma$ level is also included (solution 3).
As mentioned by U99a, and explicitly confirmed here (solution 4), the
slope for the sample with $\log P <0.4$ is significantly steeper,
confirming the result of Bauer et al. (1999).

Table~2 gives the results for FO pulsators as well, including no
cut-off in period (entry 5 as obtained by U99a, and solution 6
obtained in this paper), and using a cut-off at $\log P = 0.25$, for
reasons explained before (solutions 7 and 8).  The slope for the
shorter periods is significantly steeper, as for the FU pulsators.
The slopes for the FO pulsators are nearly identical to the ones for
the FU pulsators both above and below the cut-off.  The dispersion in
all relations is larger than the corresponding relations for the LMC,
and this is very likely due to the known depth effect of the SMC.

\begin{table*}
\caption[]{$PL$-relations in the SMC of the form $M = a \times \log P +b$}
\begin{tabular}{ccccccl} \hline
solution & $M$  &   $a$     &     $b$     & $\sigma$ & N  & Remarks \\ \hline

1& $W$  &  $-3.310 \pm 0.020$ & 16.387 $\pm$ 0.016 & 0.125 & 463 & 
                        U99a; FU; $\log P>0.4$; 2.5$\sigma$ clipping \\

2& $W$  &  $-3.300 \pm 0.021$ & 16.381 $\pm$ 0.016 & 0.126 & 446 & 
                  This paper; FU; $\log P>0.4$; 2.5$\sigma$ clipping \\

3& $W$  &  $-3.286 \pm 0.024$ & 16.366 $\pm$ 0.018 & 0.145 & 464 & 
                  This paper; FU; $\log P>0.4$; 4$\sigma$ clipping \\

4& $W$  &  $-3.539 \pm 0.058$ & 16.520 $\pm$ 0.014 & 0.142 & 743 & 
                  This paper; FU; $\log P<0.4$; 4$\sigma$ clipping \\

5& $W$  &  $-3.567 \pm 0.023$ & 15.981 $\pm$ 0.006 & 0.121 & 714 & 
                        U99a; FO; $\log P>-0.2$; 2.5$\sigma$ clipping \\

6& $W$  &  $-3.558 \pm 0.025$ & 15.978 $\pm$ 0.006 & 0.126 & 688 & 
                  This paper; FO; $\log P>-0.2$; 2.5$\sigma$ clipping \\

7& $W$  &  $-3.284 \pm 0.091$ & 15.854 $\pm$ 0.037 & 0.138 & 211 & 
                  This paper; FO; $\log P>0.25$; 3$\sigma$ clipping \\

8& $W$  &  $-3.572 \pm 0.050$ & 15.982 $\pm$ 0.067 & 0.145 & 518 & 
                  This paper; FO; $\log P<0.25$; 3$\sigma$ clipping \\

9& $W$  &  $-3.328 \pm 0.023$ & 16.395 $\pm$ 0.017 & 0.132 & 464 & 
         FU; $\log P>0.4$; 4$\sigma$ clipping; planar effect taken out \\

10&$W$  &  $-3.360 \pm 0.082$ & 15.881 $\pm$ 0.033 & 0.123 & 211 & 
         FO; $\log P>0.25$; 3$\sigma$ clipping; planar effect taken out \\

\hline
\end{tabular}
\end{table*}

To discuss the spatial structure of the SMC the FU pulsators are
considered first in determining $\theta$ and $i$ (solution 3).  An arbitrary
distance to the center of the SMC of $R = 63$ kpc is assumed, and the
center is assumed to be given by the mean of the 464 \C\ in the
sample: $\alpha_0 = 13.529$, $\delta_0 = -72.951$ degrees (J2000).
Figure~8 gives the relevant projection for $\theta = 58$ and $i = 0$
degrees.

\begin{figure}
\centerline{\psfig{figure=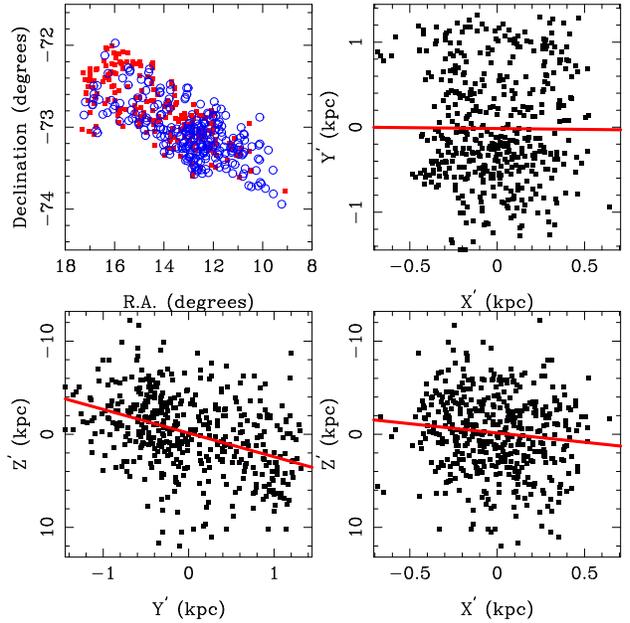,width=8.5cm}}
\caption[]{As Fig.~2 for the SMC, FU pulsators and $i = 0$ and $\theta
= 58\degr$.  }
\end{figure}

It is clear that the SMC is inclined with the NE part closer to us. A
fit with a position angle of the line-of-nodes of $\theta = 238$
($\pm$ 7), and $i = 68$ ($\pm$ 2) degrees is shown in Fig.~9. The
strong correlation in the bottom left panel is an artefact.  In the
transformation to $y^\prime$ and $z^\prime$, the dominant term is
respectively $-z_0 \sin i$ and $+z_0 \cos i$ (see the Appendix in
Weinberg \& Nikolaev, 2000). Because, as mentioned before, $z_0$ does
not correspond to the physical distance from the plane, but is
dominated by the intrinsic scatter in the $PL$-relation, $y^\prime$
and $z^\prime$ will be strongly correlated.

\begin{figure}
\centerline{\psfig{figure=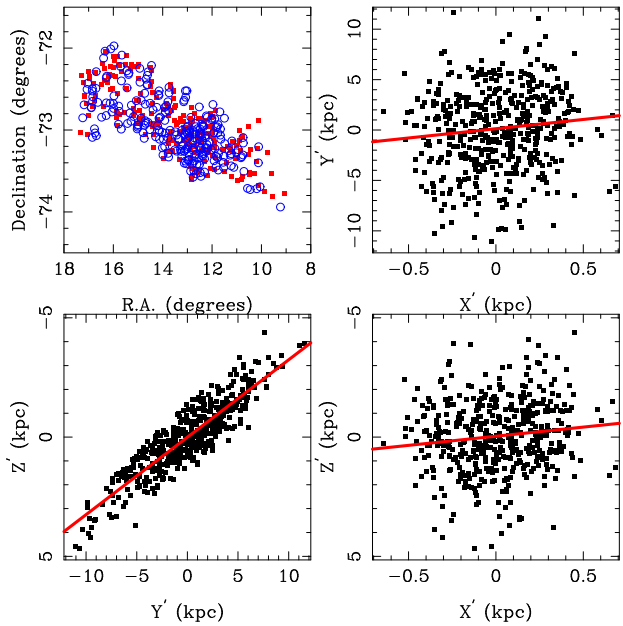,width=8.5cm}}
\caption[]{As Fig.~8 for $i = 68$ and P.A. of the line of nodes of 
$\theta = 238\degr$.
}
\end{figure}

The analysis is repeated for the FO pulsators and $\log P$ $>$0.25
(solution 7), and the result is a P.A. of the line-of-nodes 245 $\pm$
10\degr, and $i = 68 \pm 4$\degr, consistent with the results for the
FU mode pulsators.  Figure~10 shows the result. The values for the
inclination angle and the position angle of the line-of-nodes are in
good agreement with the previous results for Cepheids quoted at the
beginnning of this section, that were obtained using far fewer
Cepheids but covering a larger area.

\begin{figure}
\centerline{\psfig{figure=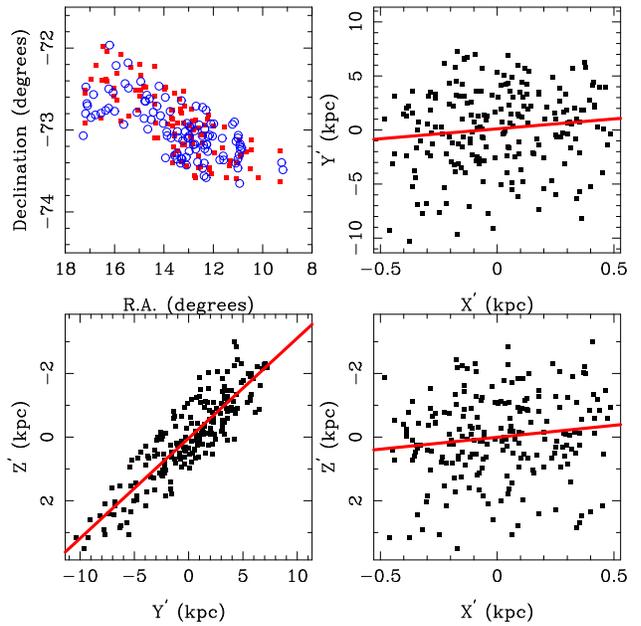,width=8.5cm}}
\caption[]{As Fig.~8 for FO pulsators, $i = 68$ and P.A. of the line
of nodes of $\theta = 240\degr$.
}
\end{figure}

The $PL$-relations are fitted taking away the depth effect, and the
results are listed in Table~2 (solutions 9\&10), and plotted in
Figs.~11-12 for FU and FO pulsators. The dispersion is lowered
significantly but is still larger than that in the $PL$-relation for
the LMC \C, indicating a substantial intrinsic depth of the SMC.

\begin{figure}
\centerline{\psfig{figure=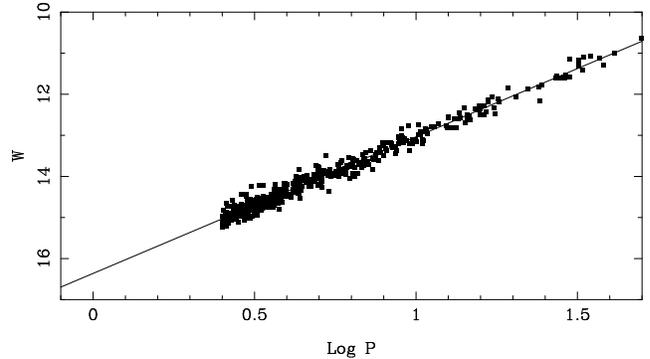,width=8.5cm}}
\caption[]{Final $PL$-relation in $W$ for FU mode pulsators with $\log P > 0.4$
in the SMC when the depth
effect has been taken out. The drawn line is the best fit.
}
\end{figure}

\begin{figure}
\centerline{\psfig{figure=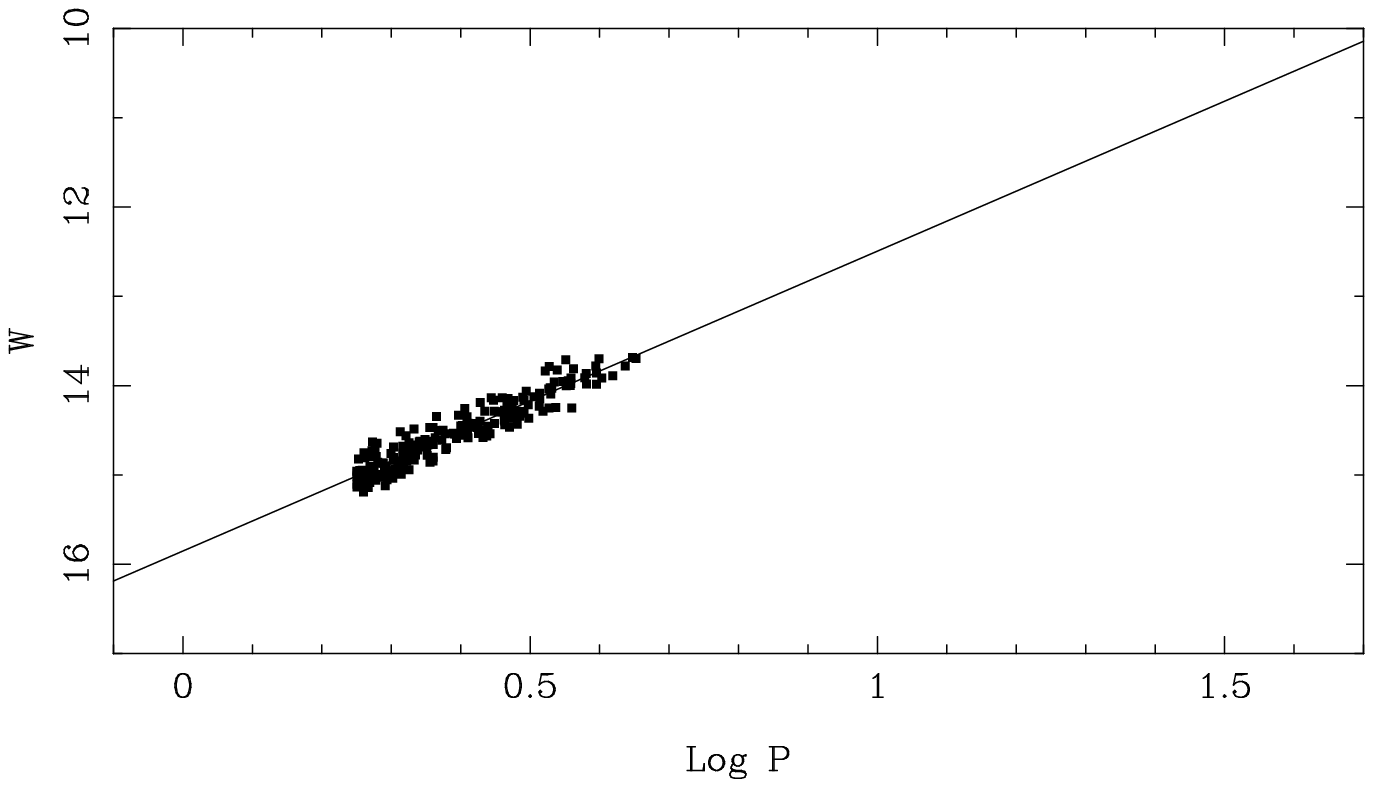,width=8.5cm}}
\caption[]{As Fig.~11 for FO mode pulsators with $\log P > 0.25$ in the SMC.
}
\end{figure}

The total depth effect, if the intrinsic scatter in the $PL$-relation
were zero, over all stars in the direction of the SMC bar is about 20
kpc (Fig.~13), with the depth along any given line-of-sight towards
the bar being about 16 kpc (see Fig.~8 lower left panel). These values
are in fact upper limits because of the intrinsic scatter in the
$PL$-relation.  Previously it was shown that the intrinsic depth of
the LMC is negligible and that the scatter in the LMC $PL$-relation is
intrinsic. If it is assumed that the intrinsic scatter in the SMC
$PL$-relation is the same, this would imply that the intrinsic depth
of the SMC translates into a dispersion of about $\sqrt{0.132^2 -
0.072^2} = 0.11$ mag meaning that the above numbers should be
multiplied by 0.7 to get the true depth effect.  The estimates for the
depth of the SMC are consistent with those based on independent
analysis, as quoted in Westerlund (1997, his Sect. 12.2.1).

\begin{figure}
\centerline{\psfig{figure=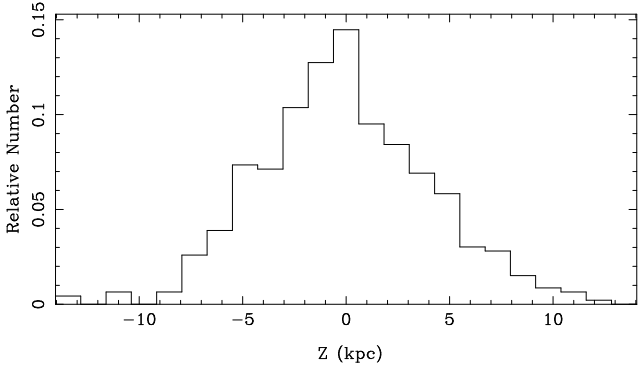,width=8.5cm}}
\caption[]{If the intrinsic spread of the $PL$-relation were zero,
this would be the distribution of stars along the line-of-sight towards the
SMC.  Positive-$z$ are towards the observer. As the intrinsic spread
is about 0.11 mag, the horizontal scale should be compressed in such a
way that the 10 kpc tick corresponds to 7 kpc.
}
\end{figure}

\begin{table}
\caption[]{Derived values for the inclination and position angle of
the line-of-nodes}
\begin{tabular}{ccccccl} \hline
Galaxy &      $i$         &     $\theta$      \\ \hline
LMC    & $18 \pm 3$\degr\ & $296 \pm 1$\degr\ \\
SMC    & $68 \pm 2$\degr\ & $238 \pm 7$\degr\ \\
\hline
\end{tabular}
\end{table}

\section{Infrared $PL$-relations}

In this section infrared $PL$-relations in the SMC and LMC will be
determined, based on the \OG\ \C\ that have a counterpart in the \M\
survey.

The observed $JHK$ magnitudes are first corrected for reddening. The
$E(B-V)$ of the respective \OG\ field is used (U99b,c), a selective
reddening $A_{\rm V}/E(B-V)$ of 3.1 is used, and the extinction curve
of Cardelli et al. (1989). In particular, $A_{\rm J} = 0.90 E(B-V)$,
$A_{\rm H} = 0.59 E(B-V)$ and $A_{\rm K} = 0.36 E(B-V)$ are used. 
Second, the \M\ magnitudes are transformed to the Carter system using
the average formulae derived in Sect.~3 (that is, 
$K$(Carter) $-$ $K$(2Mass) = +0.02, $H$(Carter) $-$ $H$(2Mass) =
$-0.02$, $J$(Carter) $-$ $J$(2Mass) = +0.05).
Third, the planar effect is taken out using the adopted values in
Table~3 based on the results obtained from the FU pulsators.

One issue that is important here is Malmquist bias. For short periods
only the on average brighter \C\ have been detected by \M\ because
the magnitudes are near the detection limit of \M. Not taking
this effect into account would lead to a derived slope of the
$PL$-relation which is too shallow (see Lanoix et al. 1999b). The
nominal completeness limits are $J$ = 15.8, $H$ = 15.1 and $K$ = 14.3
mag but at higher galactic latitude the \M\ catalog contains accurate
detections 0.5-1.0 mag fainter than this (Cutri et al. 2000).

Table~4 contains the derived $PL$-relations in the different bands, on
the Carter system, and with the planar effect taken out. Different
cut-offs in $\log P$ were tested to ensure that the slopes are
unbiased. The $PL$-relations are displayed in Figs.~14-25. 
The dispersions are generally slightly larger than quoted in the
literature (for example, 0.16, 0.12, 0.12 mag in the $PL$-relation for
FU pulsators in
respectively $J,H,K$ [Gieren et al. 1998, using 59 \C\ in the LMC]),
but this is due to the fact that single-epoch infrared photometry is
used here, while the data that Gieren et al. use is based on
intensity-mean magnitudes. Peak-to-peak amplitudes depend on period
and wavelength but can be 0.2-0.6 mag in the infrared (Laney \& Stobie
1986b). The fact that the sample studied here is an order of magnitude
larger than previously studied still allows for very accurate
determinations of the slope and zero point of the infrared $PL$-relations.

Some tests were made to check the influence of the adopted
reddening. Decreasing $E(B-V)$ in steps of 0.01 mag will make the
respective zero points in $JHK$ fainter in steps of respectively
0.009, 0.006 and 0.004 mag. For reference, based on the solutions in
the $H$-band, the average $E(B-V)$ is 0.148 for the FU pulsators, and
0.156 for the FO pulsators in the LMC (averaged over all periods), and
respectively 0.091 and 0.092 for the FU (for $\log P >0.4$) and FO (for
$\log P > 0.3$) pulsators in the SMC. 

There is some evidence that the selective reddening towards the LMC is
lower than the adopted value of 3.1 (Misselt et al. (1999) derive
values for $R$ towards LMC stars that range between 2.16 $\pm$ 0.30
and 3.31 $\pm$ 0.20, with un unweighted mean of 2.55). Chosing $R =
2.6$ would make the respective zero points in $JHK$ fainter by
respectively 0.037, 0.021 and 0.005 mag in the LMC, and fainter by
respectively 0.008, 0.008 and 0.006 mag in the SMC.

\begin{table*}
\caption[]{Infrared $PL$-relations of the form $M = a \times \log P +b$}
\begin{tabular}{ccccccl} \hline
solution & $M$  &   $a$     &     $b$     & $\sigma$ & N  & Remarks \\ \hline

1& $K_0$ &  $-3.246 \pm 0.036$ & 16.032 $\pm$ 0.025 & 0.168 & 472 & 
                  LMC; FU; $\log P > 0.4$; 3$\sigma$ clipping \\

2& $H_0$ &  $-3.236 \pm 0.033$ & 16.048 $\pm$ 0.023 & 0.161 & 493 & 
                  LMC; FU; all $\log P$; 3$\sigma$ clipping \\

3& $J_0$ &  $-3.144 \pm 0.035$ & 16.356 $\pm$ 0.025 & 0.173 & 490 & 
                  LMC; FU; all $\log P$; 3$\sigma$ clipping \\

4& $K_0$ &  $-3.381 \pm 0.076$ & 15.533 $\pm$ 0.032 & 0.137 & 238 & 
                  LMC; FO; $\log P > 0.25$; 3$\sigma$ clipping \\

5& $H_0$ &  $-3.434 \pm 0.060$ & 15.550 $\pm$ 0.024 & 0.126 & 263 & 
                  LMC; FO; all $\log P$; 3$\sigma$ clipping \\

6& $J_0$ &  $-3.299 \pm 0.066$ & 15.817 $\pm$ 0.027 & 0.140 & 267 & 
                  LMC; FO; all $\log P$; 3$\sigma$ clipping \\

7& $K_0$ &  $-3.212 \pm 0.033$ & 16.494 $\pm$ 0.026 & 0.194 & 418 & 
                  SMC; FU; $\log P > 0.4$; 3$\sigma$ clipping \\


8& $H_0$ &  $-3.160 \pm 0.032$ & 16.475 $\pm$ 0.025 & 0.182 & 414 & 
                  SMC; FU; $\log P > 0.4$; 3$\sigma$ clipping \\

9& $J_0$ &  $-3.037 \pm 0.034$ & 16.771 $\pm$ 0.027 & 0.199 & 418 & 
                  SMC; FU; $\log P > 0.4$; 3$\sigma$ clipping \\

10& $K_0$ &  $-3.102 \pm 0.155$ & 15.937 $\pm$ 0.068 & 0.178 & 156 & 
                  SMC; FO; $\log P > 0.3$; 3$\sigma$ clipping \\

11& $H_0$ &  $-3.199 \pm 0.142$ & 15.990 $\pm$ 0.062 & 0.162 & 162 & 
                  SMC; FO; $\log P > 0.3$; 3$\sigma$ clipping \\

12& $J_0$ &  $-3.104 \pm 0.159$ & 16.292 $\pm$ 0.070 & 0.183 & 158 & 
                  SMC; FO; $\log P > 0.3$; 3$\sigma$ clipping \\


\hline
\end{tabular}
\end{table*}

\begin{figure}
\centerline{\psfig{figure=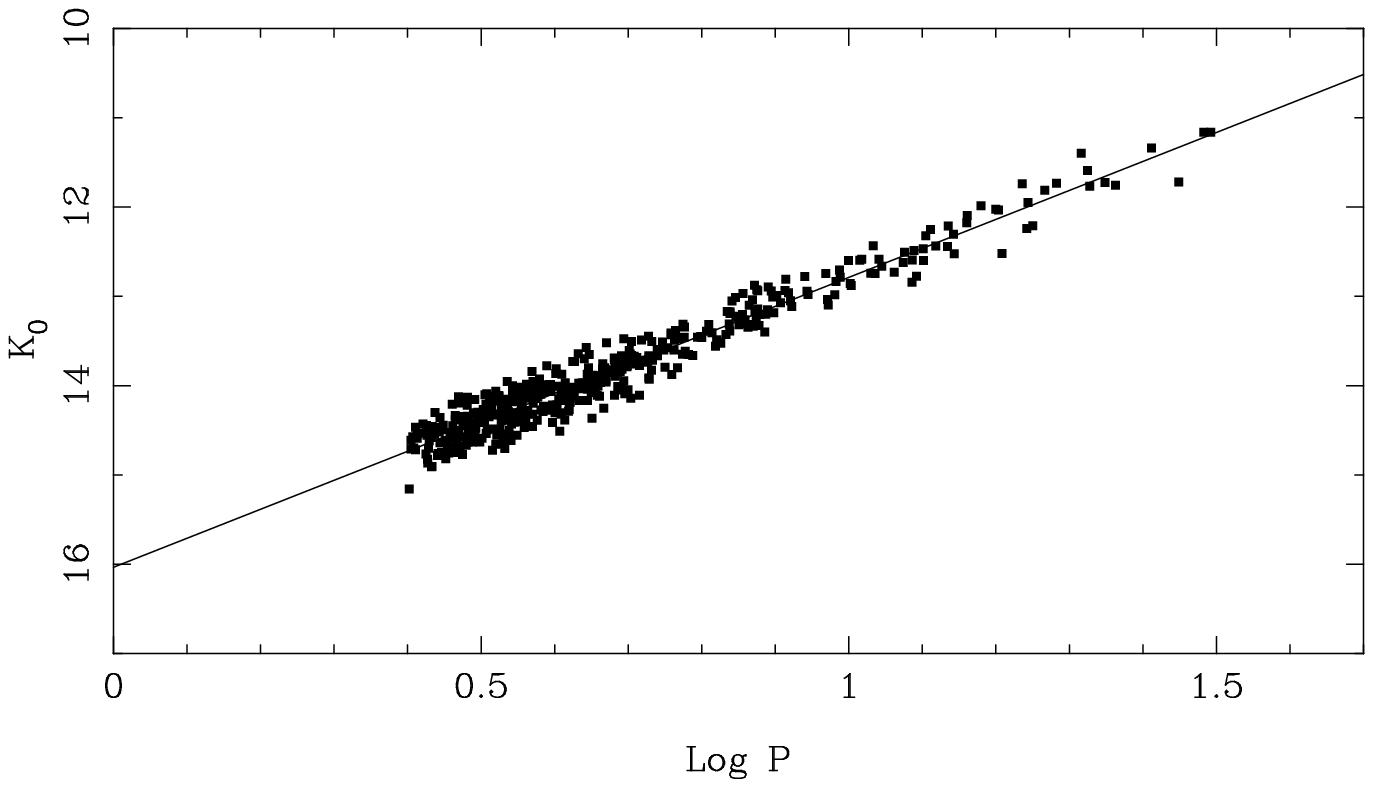,width=8.5cm}}
\caption[]{Final $PL$-relation in $K$ for FU mode pulsators in the LMC
when the depth effect has been taken out. The drawn line is the best fit.}
\end{figure}
\begin{figure}
\centerline{\psfig{figure=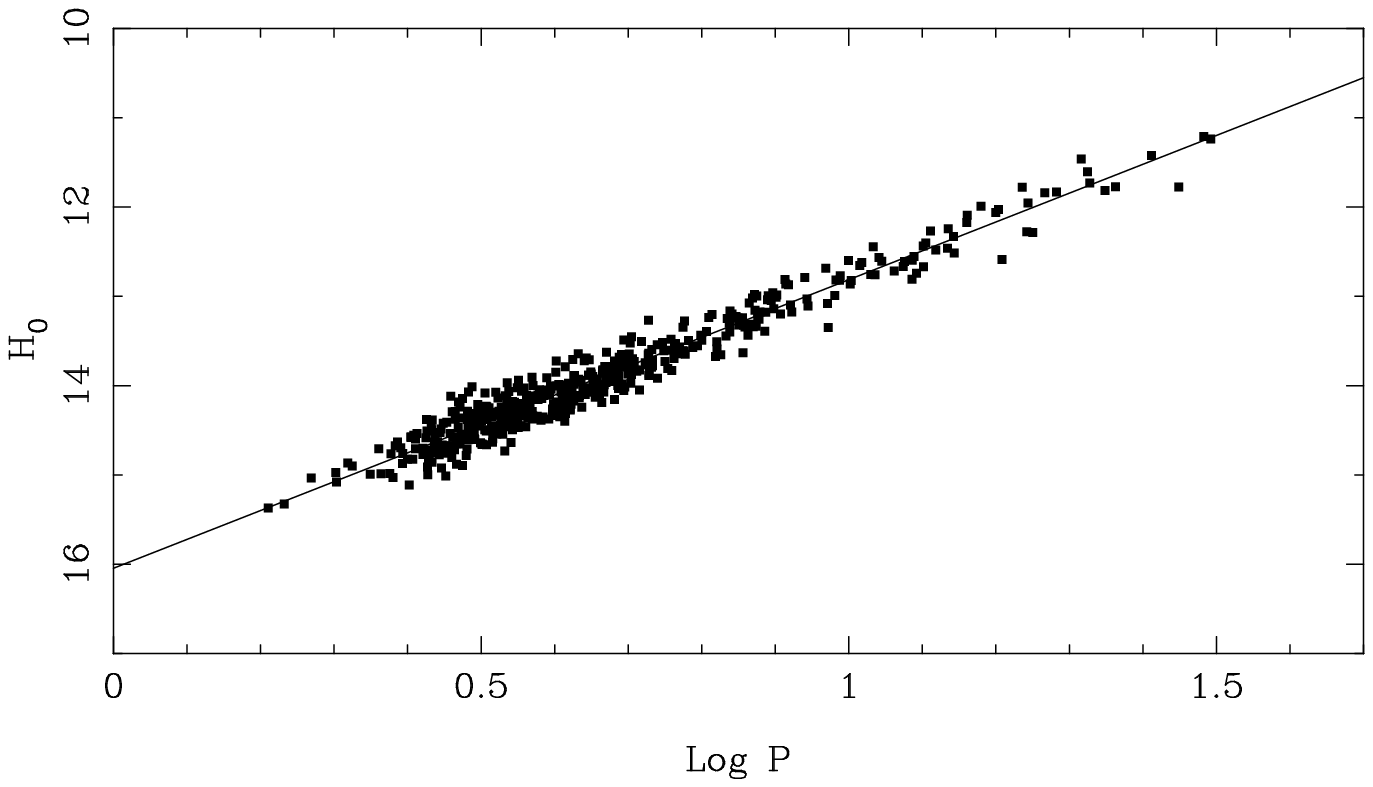,width=8.5cm}}
\caption[]{As Fig.~14: Final $PL$-relation in $H$ for FU mode pulsators in the LMC.}
\end{figure}
\begin{figure}
\centerline{\psfig{figure=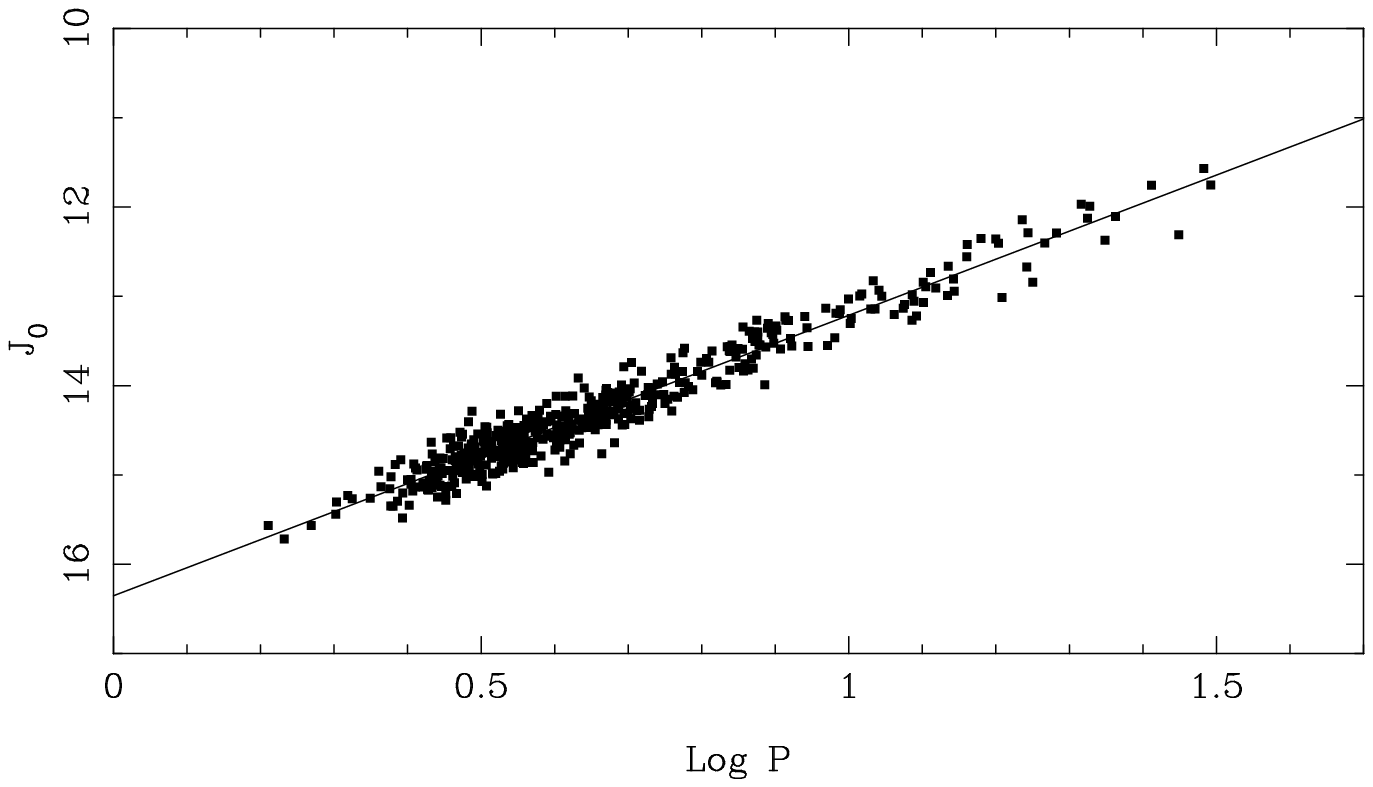,width=8.5cm}}
\caption[]{As Fig.~14: Final $PL$-relation in $J$ for FU mode pulsators in the LMC.}
\end{figure}
\begin{figure}
\centerline{\psfig{figure=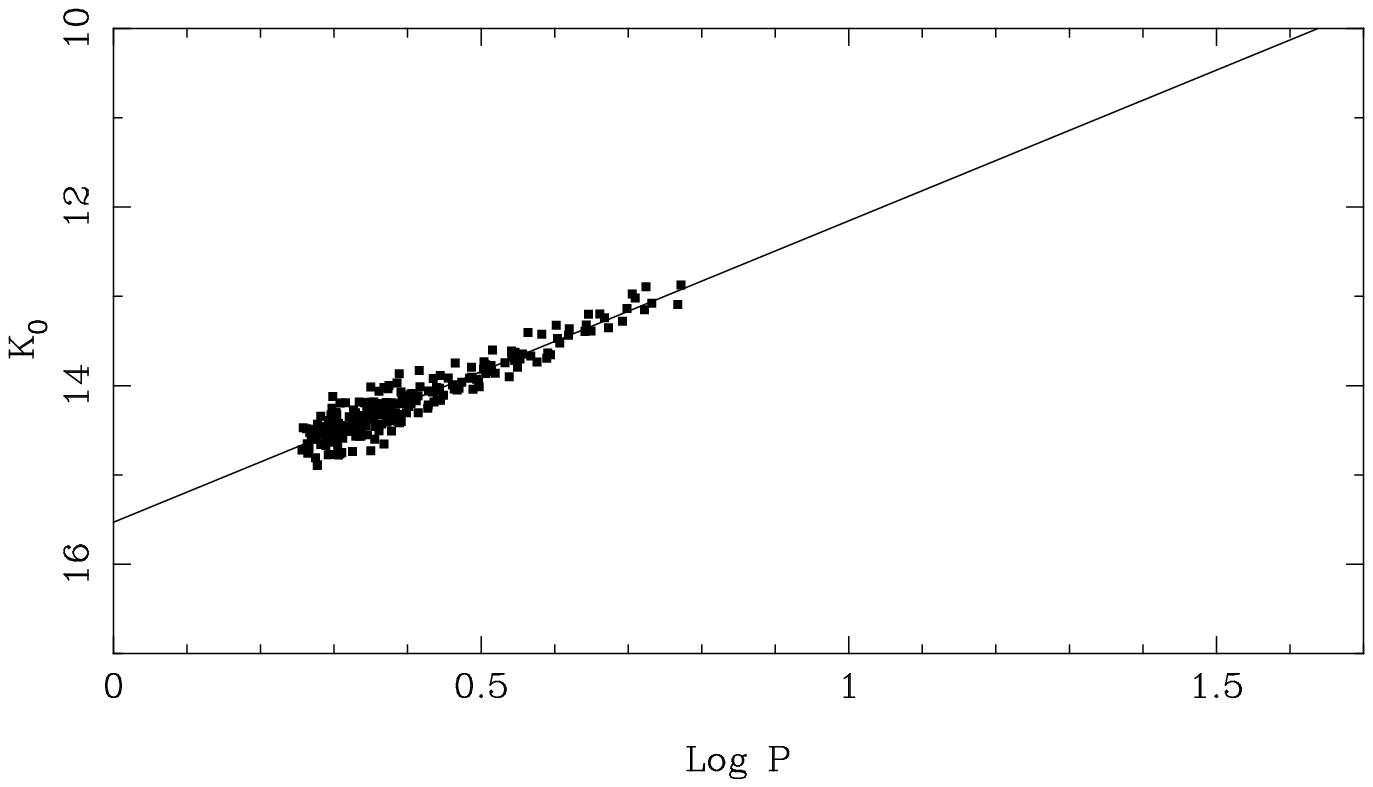,width=8.5cm}}
\caption[]{As Fig.~14: Final $PL$-relation in $K$ for FO mode pulsators in the LMC.}
\end{figure}
\begin{figure}
\centerline{\psfig{figure=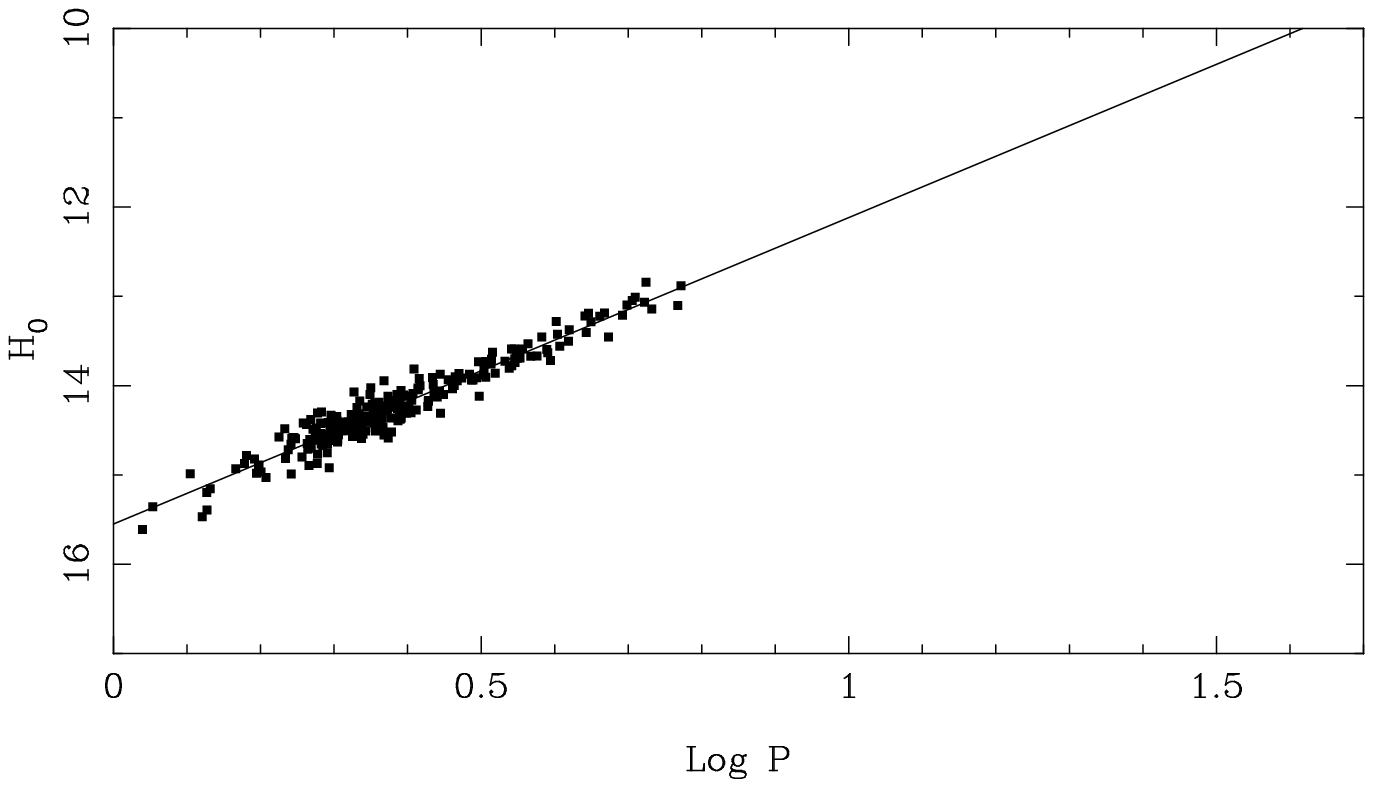,width=8.5cm}}
\caption[]{As Fig.~14: Final $PL$-relation in $H$ for FO mode pulsators in the LMC.}
\end{figure}
\begin{figure}
\centerline{\psfig{figure=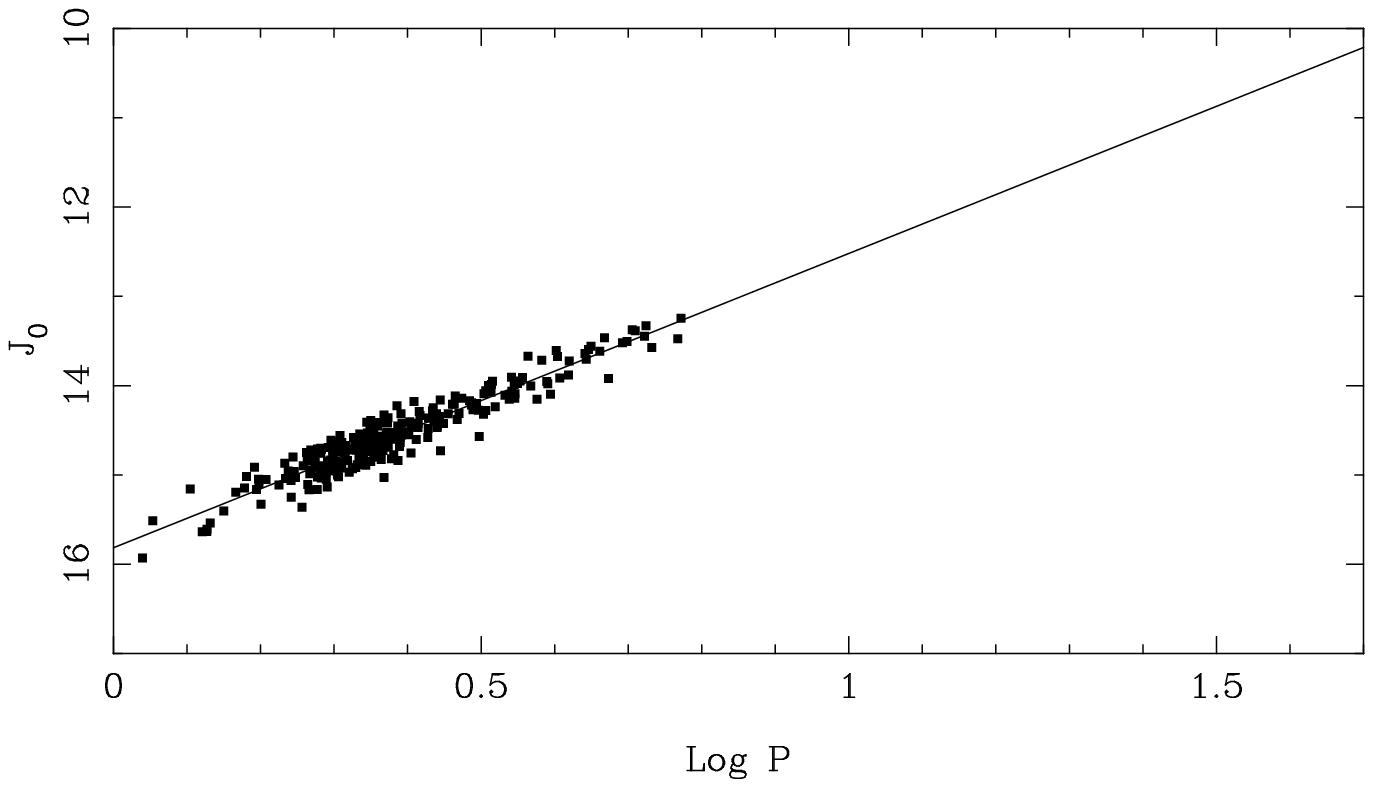,width=8.5cm}}
\caption[]{As Fig.~14: Final $PL$-relation in $J$ for FO mode pulsators in the LMC.}
\end{figure}
\begin{figure}
\centerline{\psfig{figure=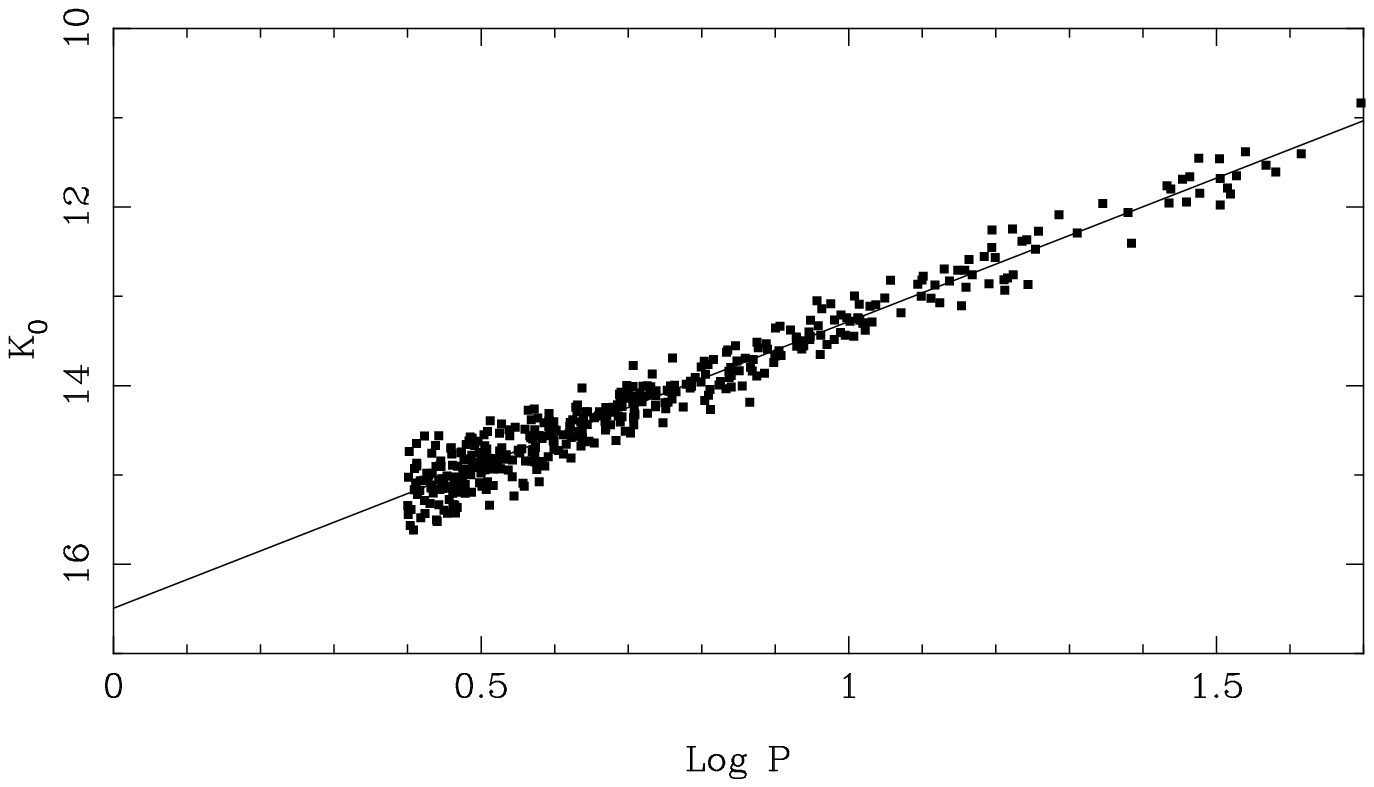,width=8.5cm}}
\caption[]{As Fig.~14: Final $PL$-relation in $K$ for FU mode pulsators in the SMC.}
\end{figure}
\begin{figure}
\centerline{\psfig{figure=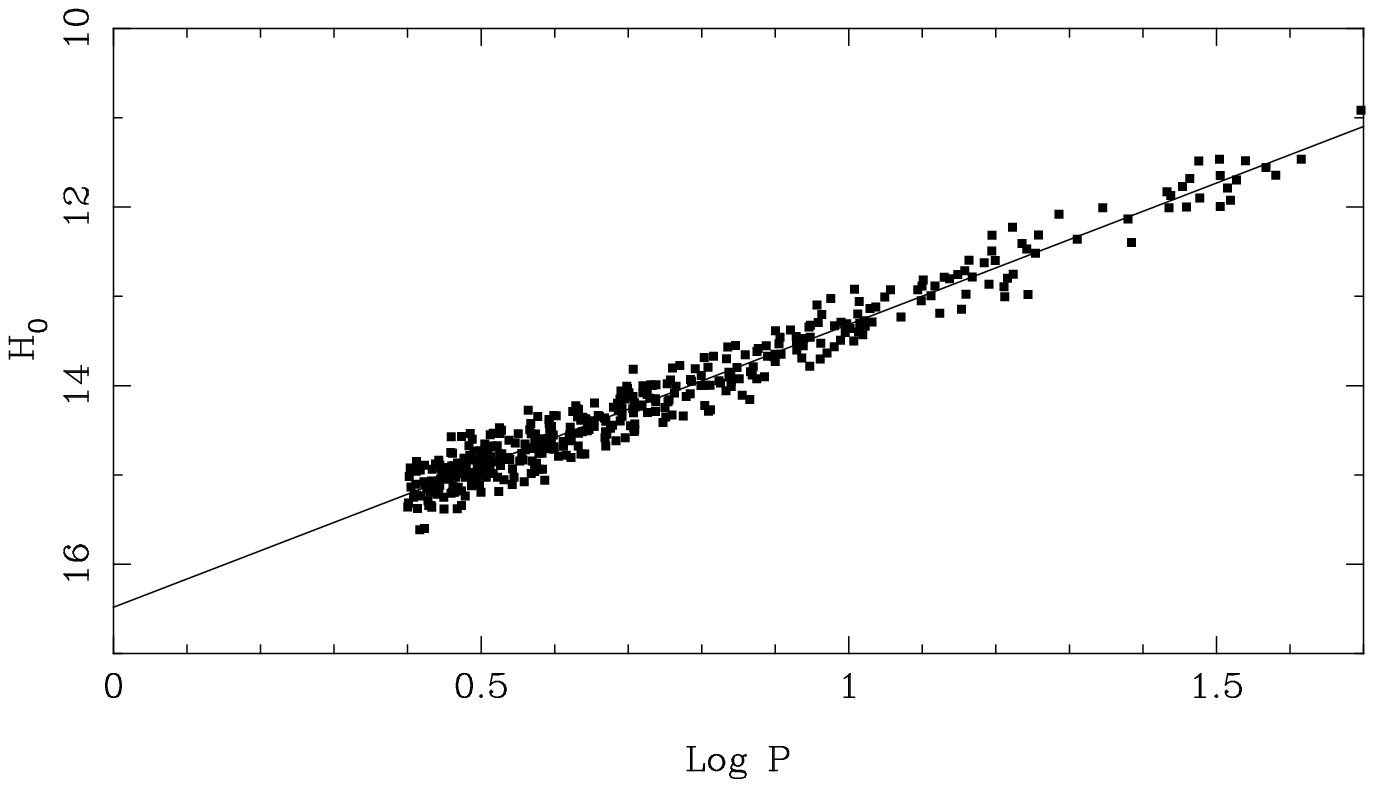,width=8.5cm}}
\caption[]{As Fig.~14: Final $PL$-relation in $H$ for FU mode pulsators in the SMC.}
\end{figure}
\begin{figure}
\centerline{\psfig{figure=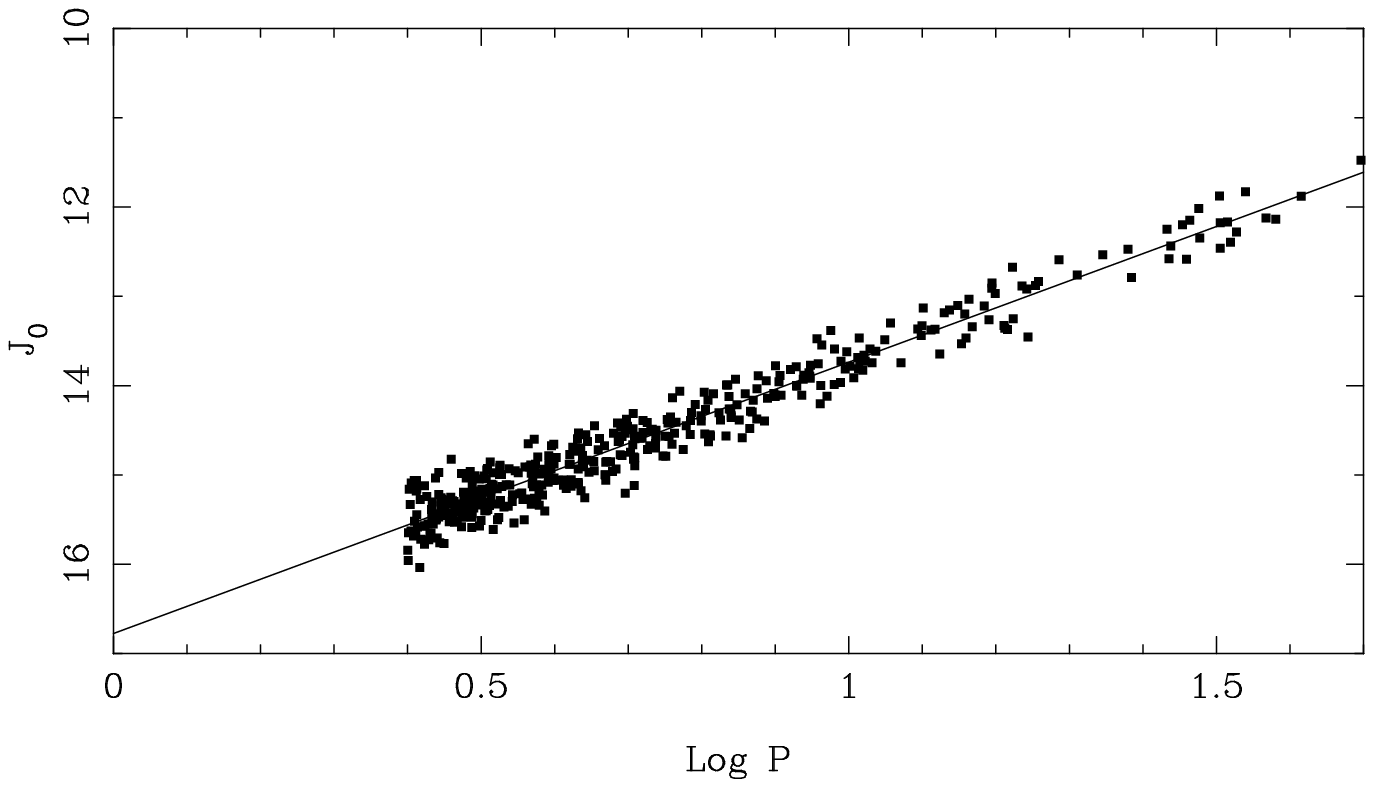,width=8.5cm}}
\caption[]{As Fig.~14: Final $PL$-relation in $J$ for FU mode pulsators in the SMC.}
\end{figure}
\begin{figure}
\centerline{\psfig{figure=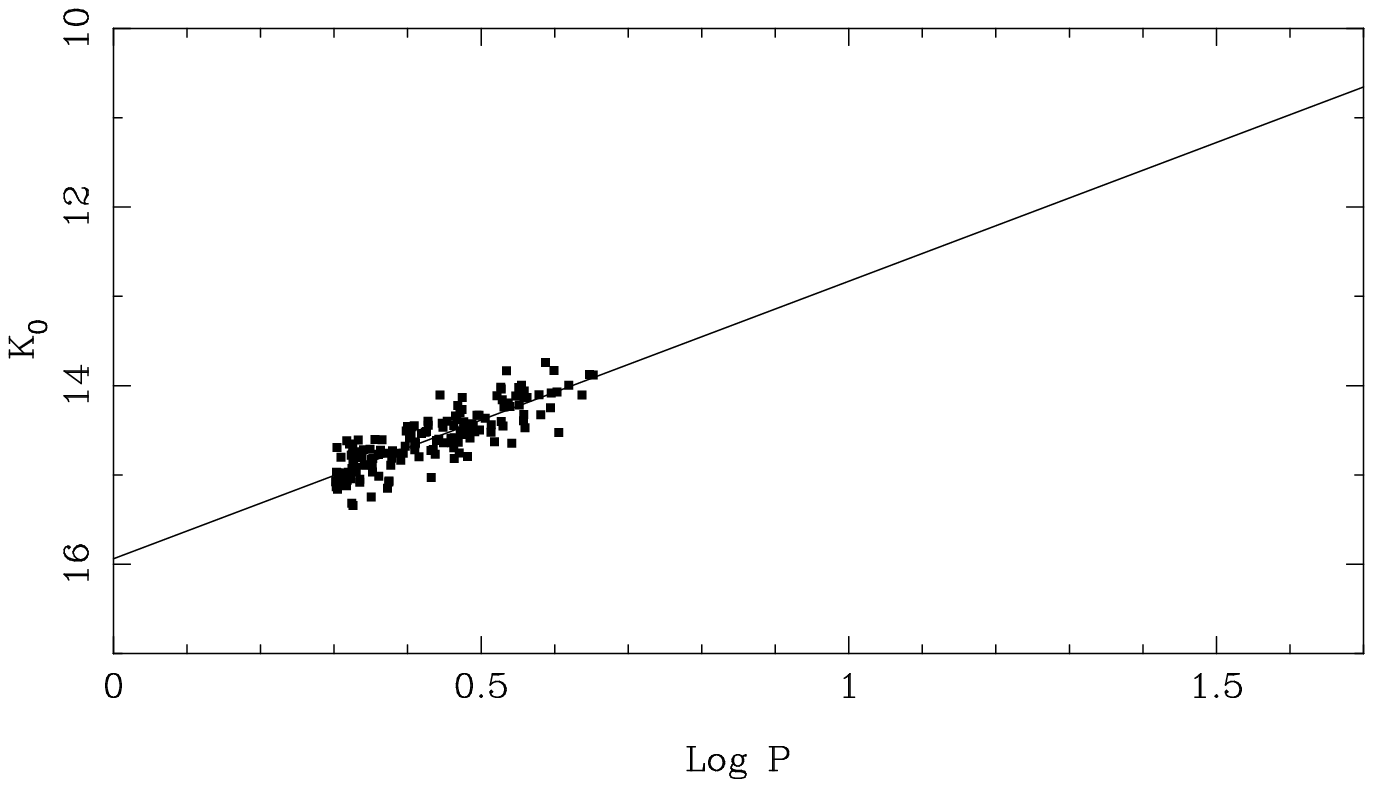,width=8.5cm}}
\caption[]{As Fig.~14: Final $PL$-relation in $K$ for FO mode pulsators in the SMC.}
\end{figure}
\begin{figure}
\centerline{\psfig{figure=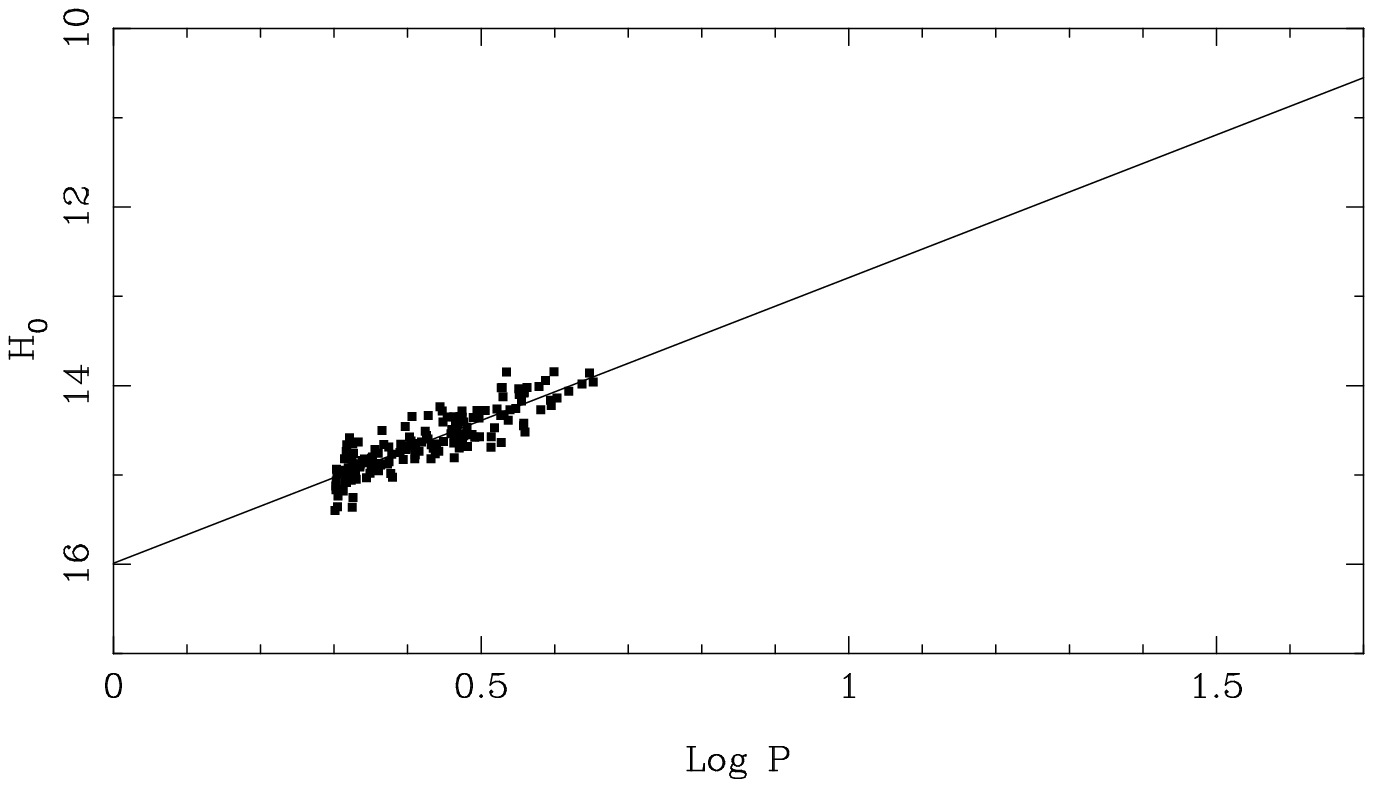,width=8.5cm}}
\caption[]{As Fig.~14: Final $PL$-relation in $H$ for FO mode pulsators in the SMC.}
\end{figure}
\begin{figure}
\centerline{\psfig{figure=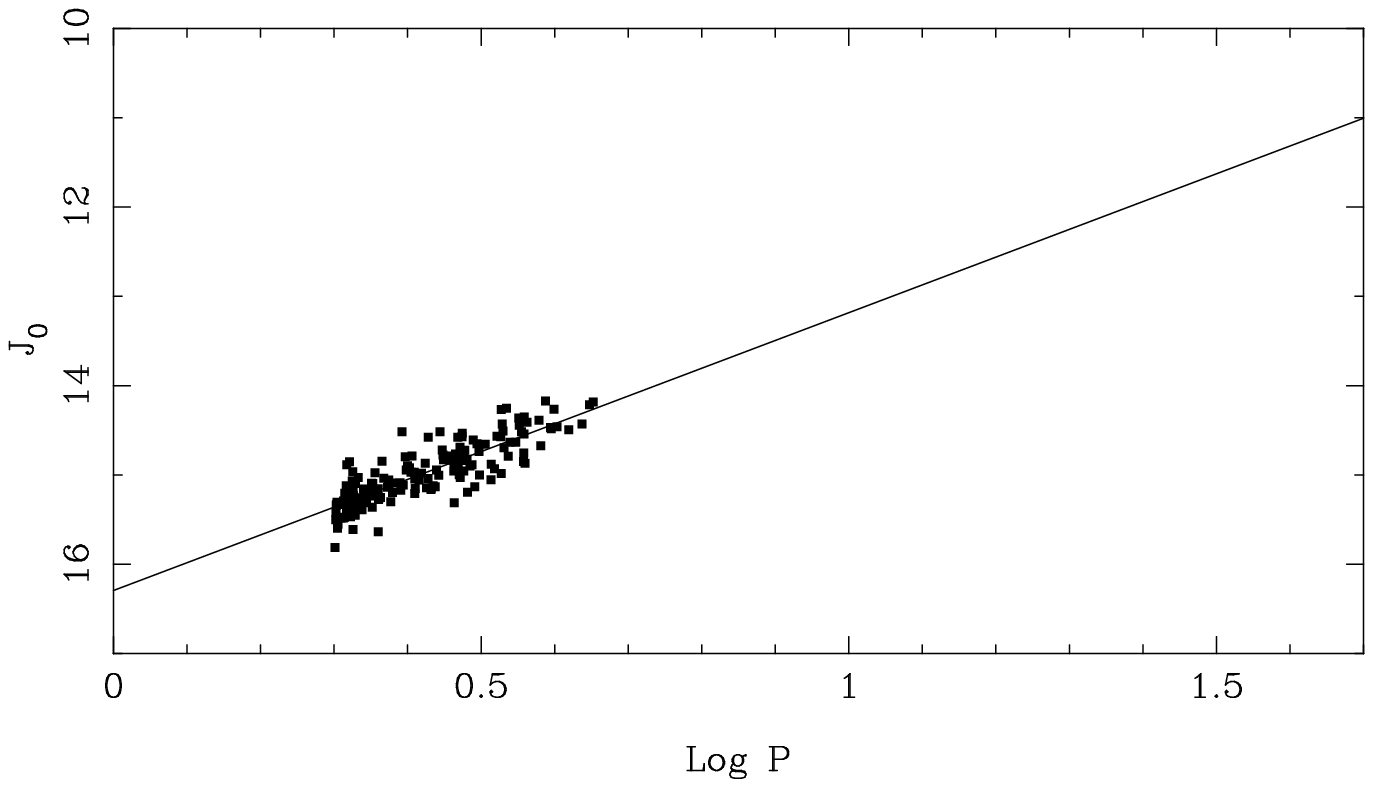,width=8.5cm}}
\caption[]{As Fig.~14: Final $PL$-relation in $J$ for FO mode pulsators in the SMC.}
\end{figure}

\section{Discussion}

\subsection{The slopes of the $PL-$relations}

One of the main results of this paper is the derivation of the slopes
of $PL$-relation in the Wesenheit-index ($W$), and in $JHK$, for both
FU and FO pulsators. In this section the slopes are intercompared,
compared to previous determinations, and to predictions of theoretical
models.

The first thing to remark is that in the LMC the slope of the
$PL$-relation of the FU and FO pulsators disagree at the
1.6-2.9$\sigma$ level. For all 4 bands the slope of the $PL$-relation
of the FO pulsators is steeper than that of the FU pulsators. In the
SMC, the slope of the $PL$-relation of the FO pulsators is steeper
than that of the FU pulsators in $WJH$, but in all 4 bands the
difference is less than 1$\sigma$, largely influenced by the fact that
the error in the slope of the $PL$-relation for the SMC FO pulsators
is relatively large.

The second thing to remark is that for the FU pulsators the slope for
the $PL$-relation of the LMC and SMC Cepheids agree within $1\sigma$
in $WK$. In $JH$ the slope of the $PL$-relation for the LMC pulsators
is steeper than that for the SMC pulsators at about the
1.7-2.2$\sigma$ level. For the FO pulsators the slope of the
$PL$-relation for the LMC \C\ is steeper than that of the SMC \C\ in
all 4 bands but the difference is 1.5$\sigma$ in $HK$ and less in $WJ$.

Table~5 collects slopes quoted in the literature both from
observations and theoretical models, and the final values derived in
the present paper.  Regarding the Wesenheit-index, the only previous
determination was by Tanvir (1999), who used a different definition
for this index and so can not be compared directly to the present
result. The results by U99a are essentially based on the same sample
considered here and hence are very close to the present results.

Regarding the IR relations, the slopes derived in the present work are
the most accurate available for the moment. The slopes agree with in
the errors with all previous determinations. The previously most
accurate slopes were by Gieren et al. (1998)\footnote{I have
calculated the error in the slope using the data they kindly provided
(their Table~9), and made a division on period as the sample
they consider almost exclusively contains \C\ with long periods,
unlike the \OG\ sample.}. Also here the agreement is excellent. The
reliability of the result is enhanced if one considers that the Gieren
et al. sample contains mainly long-period \C\ while the \OG\ sample is
dominated by short-period \C.

The agreement with theoretically predicted slopes is less
satisfactory. For the Wesenheit-index the predicted slopes by Bono \&
Marconi (2000, private communicataion) are too shallow compared to the
observations at the 2-4$\sigma$ level. In $J$ the models by Alibert et
al. (1999) are clearly too steep. In $K$ the agreement with the Bono
et al. (1999) models is within the errors, but the Alibert et
al. (1999) models predict slopes that are too steep.

Bono et al. (1999) are the only ones to predict the slopes for
FO-pulsators (averaged over $Z$ = 0.02, 0.008 and 0.004). Their
prediction is in good agreement with the observed slope in the LMC and
within 2$\sigma$ of the observed slope in the SMC.

Clearly, the present results can be used as tight constraint for
theoretical modelling as for the first time accurate slopes in the
infrared for both FU and FO pulsators are presented.

\begin{table*}
\caption[]{Slopes of the $PL$-relations.}

\small

\begin{tabular}{ccll} \hline
Slope         & Colour & System & Reference \\ \hline

\multicolumn{4}{c}{Emperical results, FU pulsators}  \\

$-3.337 \pm 0.013$ & Wesenheit & LMC & this work \\
$-3.411 \pm 0.036$ & Wesenheit & LMC & Tanvir (1999); for $W = I - 1.45\;(V-I)$ \\
$-3.300 \pm 0.011$ & Wesenheit & LMC & Udalski et al. (1999a) \\
$-3.328 \pm 0.023$ & Wesenheit & SMC & this work \\
$-3.310 \pm 0.020$ & Wesenheit & SMC & Udalski et al. (1999a) \\


$-3.144 \pm 0.035$ & J & LMC & this work \\
$-3.129 \pm 0.052$ & J & LMC & Gieren et al. (1998) \\
$-3.147 \pm 0.065$ & J & LMC & this work; 54 stars with $\log P > 0.833$ from Gieren et al. (1998)\\
 $-2.99 \pm 0.59$  & J & LMC & this work;  5 stars with $\log P < 0.833$ from Gieren et al. (1998)\\
 $-3.31 \pm 0.10$  & J & LMC & Madore \& Freedman (1991) \\
 $-3.31 \pm 0.12$  & J & LMC & Laney \& Stobie (1986a; their solution 1b)  \\

$-3.037 \pm 0.034$ & J & SMC & this work \\
 $-3.22 \pm 0.17$  & J & SMC & Laney \& Stobie (1986a; their solution 10b)  \\


$-3.236 \pm 0.033$ & H & LMC & this work \\
$-3.249 \pm 0.044$ & H & LMC & Gieren et al. (1998) \\
$-3.246 \pm 0.056$ & H & LMC & this work; 54 stars with $\log P > 0.833$ from Gieren et al. (1998)\\
 $-3.49 \pm 0.26$  & H & LMC & this work;  5 stars with $\log P < 0.833$ from Gieren et al. (1998)\\
 $-3.72 \pm 0.07$  & H & LMC & Caldwell \& Laney (1991)  \\
 $-3.40 \pm 0.10$  & H & LMC & Laney \& Stobie (1986a; their solution 2b)  \\
 $-3.37 \pm 0.10$  & H & LMC & Madore \& Freedman (1991) \\

$-3.160 \pm 0.032$ & H & SMC & this work \\
 $-3.25 \pm 0.07$  & H & SMC & Caldwell \& Laney (1991)  \\
 $-3.36 \pm 0.15$  & H & SMC & Laney \& Stobie (1986a; their solution 11b)  \\


$-3.246 \pm 0.036$ & K & LMC & this work \\
$-3.267 \pm 0.041$ & K & LMC & Gieren et al. (1998) \\
$-3.304 \pm 0.052$ & K & LMC & GO00 [54 stars with $\log P > 0.833$ from Gieren et al. (1998)]\\
 $-3.37 \pm 0.39$  & K & LMC & GO00 [ 5 stars with $\log P < 0.833$ from Gieren et al. (1998)]\\
 $-3.42 \pm 0.09$  & K & LMC & Madore \& Freedman (1991) \\
 $-3.41 \pm 0.10$  & K & LMC & Laney \& Stobie (1986a; their solution 3b)  \\

$-3.212 \pm 0.033$ & K & SMC & this work \\
 $-3.38 \pm 0.15$  & K & SMC & Laney \& Stobie (1986a; their solution 12b)  \\

\multicolumn{4}{c}{Theoretical results, FU pulsators}  \\

$-3.02 \pm 0.04$   & Wesenheit & 0.02  & Bono \& Marconi (2000, private communication) \\
$-3.17 \pm 0.04$   & Wesenheit & 0.008 & Bono \& Marconi (2000, private communication) \\
$-3.21 \pm 0.05$   & Wesenheit & 0.004 & Bono \& Marconi (2000, private communication) \\

 $-3.286$          & J & 0.01  & Alibert et al. (1999)\\
 $-3.262$          & J & 0.004 & Alibert et al. (1999)\\

 $-3.19 \pm 0.09$  & K & 0.008 & Bono et al. (1999) \\
 $-3.395$          & K & 0.01  & Alibert et al. (1999) \\
 $-3.27 \pm 0.09$  & K & 0.004 & Bono et al. (1999)  \\
 $-3.369$          & K & 0.004 & Alibert et al. (1999) \\

\multicolumn{4}{c}{Emperical and theoretical results, FO pulsators}  \\

$-3.381 \pm 0.076$ & K & LMC & this work, FO-pulsators \\
$-3.102 \pm 0.155$ & K & SMC & this work, FO-pulsators \\
 $-3.44 \pm 0.05$  & K & all $Z$ & Bono et al. (1999), FO-pulsators \\

\hline
\end{tabular}
\end{table*}

\subsection{The relative distance SMC-LMC}

These results allow a determination of the relative distance SMC to
LMC, based on the various $PL$-relations. Since the slopes of the
$PL$-relations in SMC and LMC are not exactly the same, the absolute
magnitudes are calculated for $\log P = 0.5$ when using FU pulsators,
and $\log P = 0.3$ for FO pulsators for each Cloud and then
subtracted, using the $PL$-relations where the distance effect has
been taken out. The results are listed in Table~6. The errors are
based on the errors in the slope and zero point. It seems that the
difference in distance modulus is somewhat less using the IR
$PL$-relations than from the Wesenheit-index. A mean using all
available 8 determinations yields ${\Delta}_{\rm SMC-LMC} = 0.50 \pm
0.02$ mag.  This is in good agreement with the original determination
in U99c based primarily on the Wesenheit index for the FU pulsators,
and the 0.50 $\pm$ 0.03 determined by Cioni et al. (2000b) as the
average of the difference in $JHK$ and $m_{\rm bol}$ of the TRGB
(Tip of the Red Giant Branch) magnitudes based on \DE\ data for the
two Clouds.

All this assumes no explicit metallicity correction. As recapitulated
in GO00 there is no agreement between theory and empirical evidence,
and amongst different theoretical models, about the extent of a
metallicity dependence of the $PL$-relation, if any. In the infrared
and Wesenheit-index there are no empirical estimates for the
metallicity dependence. Table~7 summarises the predictions of $\Delta
M = M$(LMC) $-M$(SMC) in $W,J,K$ for 2 sets of recent models at $\log P =
0.5$. Again the disagreement is obvious and at a level of upto 0.1 mag
which is not negligible.

\begin{table}
\caption[]{The relative distance modulus SMC-LMC}
\begin{tabular}{cc} \hline
$\Delta$        & Remarks \\ \hline
0.51 $\pm$ 0.03 & U99c; W; FU \\

0.510 $\pm$ 0.023 & present work; W; FU \\
0.522 $\pm$ 0.042 & present work; W; FO \\

0.479 $\pm$ 0.044 & present work; K; FU \\
0.488 $\pm$ 0.091 & present work; K; FO \\
0.465 $\pm$ 0.041 & present work; H; FU \\
0.511 $\pm$ 0.081 & present work; H; FO \\
0.469 $\pm$ 0.044 & present work; J; FU \\
0.534 $\pm$ 0.092 & present work; J; FO \\

0.475 $\pm$ 0.022 & mean of the 6 IR determinations \\
0.504 $\pm$ 0.015 & mean of all 8 determinations \\

\hline
\end{tabular}
\end{table}

\begin{table}
\caption[]{Metallicity dependence of the absolute magnitude between
LMC and SMC, $\Delta M = M$(LMC) $-M$(SMC) at $\log P = 0.5$ for FU pulsators.}
\begin{tabular}{rrrl} \hline
${\Delta M}_{\rm W}$ & ${\Delta M}_{\rm J}$ & ${\Delta M}_{\rm K}$
& Reference \\ \hline
 $+0.030$ & $-$      & $+0.020$ & Bono et al. (1999)\\
 $-$      & $-0.093$ & $-0.091$ & Alibert et al. (1999)\\
\hline
\end{tabular}
\end{table}

\section{The zero point of the Galactic $PL$-relation}

GO00 discuss in detail the sample of Galactic \C\ with \HP\ parallaxes
and the methods and relations used to compute the zero point of the
Galactic $PL$-relation. Since the slopes in $W$ and $K$ differ from
those used in GO00, new zero points calculations are necessary.  They
are summarised in Table~8, and have the same format as in GO00.  It
was remarked in GO00, and repeated here, that the zero point of both
the $V$ and $I$ $PL$-relations for the sample of the 63 stars with
infrared data differs by 0.10 mag from those using the full
samples. Until infrared date for more \C\ in \HP\ becomes available to
better constrain this zero point, it seems prudent to add 0.10 mag to
the zero points in $K$ as listed in Table~8.

As discussed in GO00, when taking into account the errors in
reddening, period-colour relations, etc., an error term of 0.028 for
$W$ and 0.011 in $K$ should be added in quadrature to the quoted
errors, and the zero points should also be corrected for
Malmquist-bias (+0.01 mag) as derived in GO00 from theoretical
calculations and numerical simulations. The final zero points for the
4 values listed in Table~8 become respectively $-2.711 \pm 0.108$,
$-2.718 \pm 0.108$, $-2.517 \pm 0.169$ and $-2.545 \pm 0.169$.

These zero points can now be combined with the zero points of the LMC
and SMC $PL$-relations to yields distances of 18.60 $\pm$ 0.11 ($W$)
and 18.55 $\pm$ 0.17 ($K$) for the LMC, and 19.11 $\pm$ 0.11 ($W$) and
19.04 $\pm$ 0.17 ($K$) for the SMC. The distance to the LMC based on
the Wesenheit index is identical to the value derived in GO00, which
was based on the results of Tanvir (1999) for the Wesenheit index for
a much smaller sample of LMC Cepheids. To these distances, systematic
uncertainties must be considered due to the fact that the slope of the
galactic $PL$-relation may not be the same as in the MCs and the
uncertain metallicity effect. In GO00 these unceratinties were
estimated to be $(\pm 0.08 \;{\rm slope})(^{+0.08}_{-0.15} \;{\rm
metallicity})$ in $W$, and $(\pm 0.03 \;{\rm slope})$ $(\pm 0.06
\;{\rm metallicity})$ in $K$.

I do not want to enter here into a discussion about the ``long'' and
``short'' distance scale to the LMC (see Gibson 1999 for an overview).
The latest ``short'' distance modulus to the LMC based on the Red
Clump method stands at 18.27 $\pm$ 0.07 (Popowski 2000) or 18.24 $\pm$
0.08 (Udalski 2000). That to the SMC at 18.77 $\pm$ 0.08 (Popowski 2000).
The distance to the LMC and SMC derived here are ``long'' and in excellent
agreement with those derived recently by Cioni et al. (2000b) using
the TRGB-method from \DE\ data of respectively 18.55 $\pm$ 0.04
(formal) $\pm$ 0.08 (systematic) and 18.99 $\pm$ 0.03 (formal) $\pm$
0.08 (systematic), and with the distance to the SMC of 19.05 $\pm$
0.017 (standard error) $\pm$ 0.043 (systematic error) using
FU/FO and FO/SO (Second Overtone) double-mode Cepheids 
(Kov\'acs 2000) and 19.11 $\pm$ 0.08 using pure 
SO and FO/SO double-mode Cepheids (Bono et al. 2000).


\begin{table*}
\caption[]{Values for the zero point for Galactic $PL$-relations}
\begin{tabular}{rrrcrl} \hline
Colour& N &   Zero point     & Assumed  &  Total  & Remarks \\
      &   &                  & Slope    & Weight  &         \\ 
\hline
W & 191 & $-2.721 \pm$ 0.104 & $-3.337$ & 5295.5  & All stars \\
W & 191 & $-2.728 \pm$ 0.104 & $-3.328$ & 5331.0  & All stars \\
K &  63 & $-2.627 \pm$ 0.169 & $-3.246$ & 1855.5  & All stars \\
K &  63 & $-2.655 \pm$ 0.169 & $-3.212$ & 1905.7  & All stars \\
\hline
\end{tabular}
\end{table*}

\subsection*{Acknowledgements}
This publication makes use of data products from the Two Micron All
Sky Survey, which is a joint project of the University of
Massachusetts and the Infrared Processing and Analysis
Center/California Institute of Technology, funded by the National
Aeronautics and Space Administration and the National Science
Foundation.

It is a pleasure to thanks Ren\'e Oudmaijer and Maurizio Salaris for
comments on earlier versions of this paper, and Giuseppe Bono and
Marcella Marconi for calculating the $PL$-relation in the
Wesenheit-index. Pascal Fouqu\'e is thanked for providing tabular
material in Gieren et al. (1998) in electronic format.

This research has made use of the SIMBAD database, operated at CDS,
Strasbourg, France.

\section*{Appendix A}

In this Appendix a comparison is presented between the \DE\ $I$ and
\OG\ $I$ and \DE\ $JK$ and \M\ $JK$ for the respectively 173 and 141
LMC and SMC \C\ in common. This has its limitations since the
comparison is done using variable stars. Yet it may still be of
interest to other workers, as such a comparison has not been done
yet. The variability would increase the spread in any correlation
but should not wash out any colour terms, if they exist.

The observed $IJK$ magnitudes are first corrected for reddening. The
$E(B-V)$ of the respective \OG\ field is used (U99b,c), a selective
reddening $A_{\rm V}/E(B-V)$ of 3.1 is used, and the extinction curve
of Cardelli et al. (1989). In particular, $A_{\rm I} = 1.87 E(B-V)$,
$A_{\rm J} = 0.90 E(B-V)$ and $A_{\rm K} = 0.36 E(B-V)$ are
used. Since the effective wavelengths of the \DE\ and \OG\ $I$ and the
\DE\ and \M\ $JK$ are very similar, no distinction between the
respective filtersets is made. At the present level of accuracy this
effect of differential reddening between the slightly different
filters should be entirely negligible.

Figure~A1  shows the differences between \M\ and \DE\ $J_0$ and $K_0$
plotted versus \M\ $(J-K)_0$, and \OG\ and \DE\ $I_0$ and plotted
versus \DE\ $(I-J)_0$. In the left panel, all stars are plotted, in
the right panel a sub-sample is plotted using a criterium on the total
error. Linear least-square fits were made and the results are listed
in Table~A1.

From this data, there is no evidence for a colour-term in $I$ and $J$
and the difference between \OG\ and \DE\ $I_0$, and \M\ and \DE\ $J_0$
is less than 0.10 magnitude. In $K$ there seems to be a color term at
the 3$\sigma$ level when all stars are used, which still is present at
the 2$\sigma$ level when a subsample of stars with the smaller
photometric errors is being used. The average difference
$K_0$(2Mass) $-$ $K_0$(Denis) is +0.20 magnitudes, which is comparable to
the observed scatter, which is an upper limit to the intrinsic scatter
as two single-epoch data are being compared.

The conclusion is drawn from the present dataset that there is no
evidence for colour terms when comparing \OG\ and \DE\ $I_0$, and \M\
and \DE\ $J_0$, and that any difference between the systems is less
than 0.10 magnitude. Regarding \M\ and \DE\ $K_0$, there is evidence
for a colour term, which amounts to a difference $K_0$(2Mass) $-$
$K_0$(Denis) = +0.20 magnitudes for a typical colour of $(J-K)_0 = 0.4$.

\setcounter{table}{0}
\renewcommand{\thetable}{A\arabic{table}}
\setcounter{figure}{0}
\renewcommand{\thefigure}{A\arabic{table}}

\begin{table*}
\caption[]{Comparing photometric systems. Fits of the form $y = a \times
x + b$}
\begin{tabular}{crccccc} \hline
         $y$             &  $a$             & $x$                 & 
   $b$            & $\sigma$ & N \\ \hline
$K_0$(2Mass)-$K_0$(Denis) & $-0.62 \pm 0.17$ & $(J-K)_0$ (2Mass) & 
$0.48 \pm 0.07$ & 0.24 & 141 \\
$K_0$(2Mass)-$K_0$(Denis) & $-0.44 \pm 0.17$ & $(J-K)_0$ (2Mass) & 
$0.35 \pm 0.08$ & 0.20 & 97 \\
$J_0$(2Mass)-$J_0$(Denis) & $0.03 \pm 0.12$ & $(J-K)_0$ (2Mass) & 
$0.09 \pm 0.05$ & 0.17 & 141 \\
$J_0$(2Mass)-$J_0$(Denis) & $0.15 \pm 0.15$ & $(J-K)_0$ (2Mass) & 
$0.03 \pm 0.07$ & 0.18 & 110 \\
$I_0$(OGLE)-$I_0$(Denis) & $-0.11 \pm 0.15$ & $(I-J)_0$ (DENIS) & 
$0.14 \pm 0.09$ & 0.33 & 173 \\
$I_0$(OGLE)-$I_0$(Denis) & $0.10 \pm 0.23$ & $(I-J)_0$ (DENIS) & 
$0.03 \pm 0.14$ & 0.32 & 119 \\
\hline
\end{tabular}
\end{table*}

\begin{figure*}
\centerline{\psfig{figure=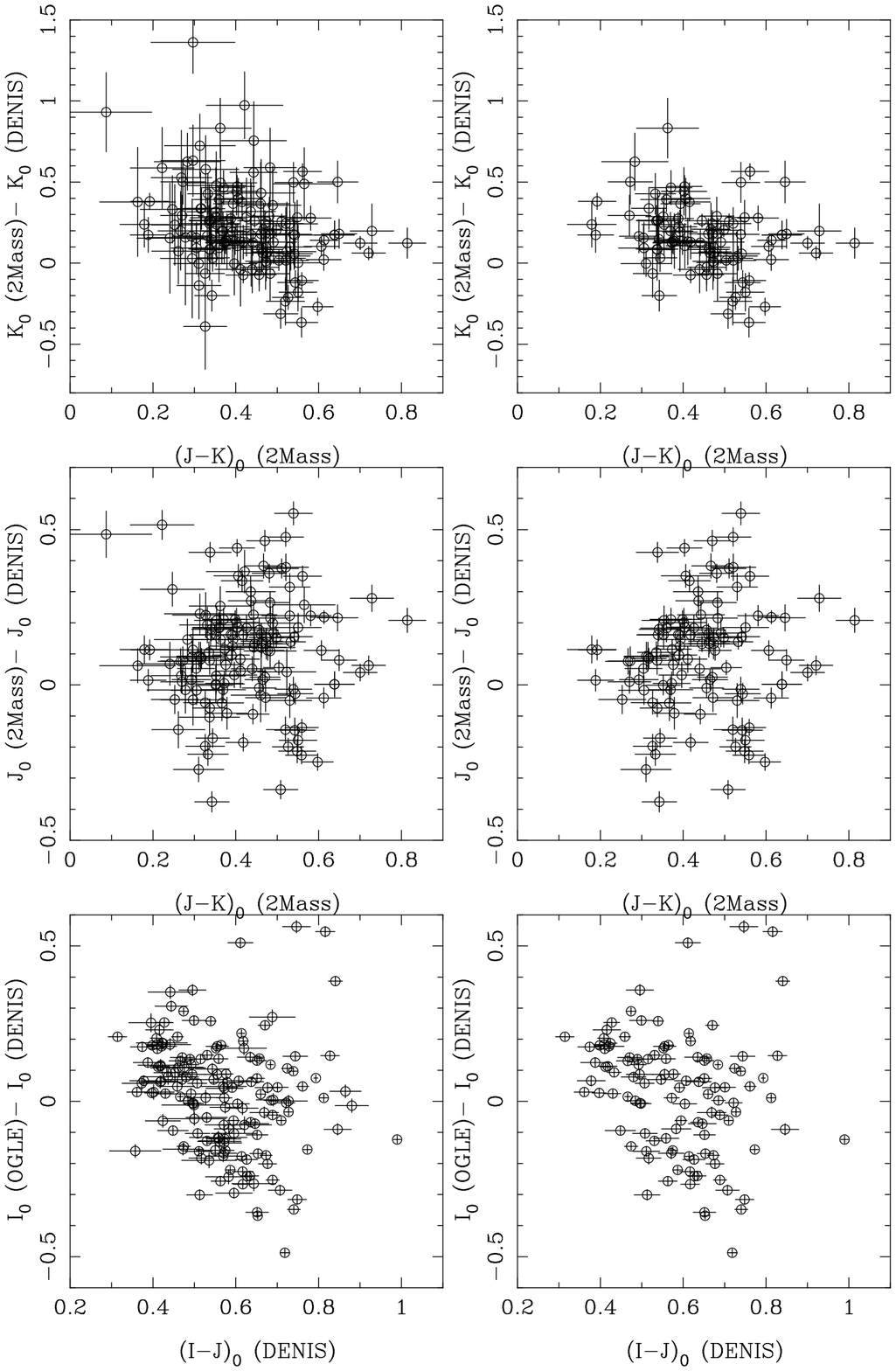,width=10.5cm}}
\caption[]{Comparison of \DE\ $IJK$ with \OG\ $I$ and \M\ $JK$. The
left hand panel contains all sources, for the right hand a selection
on the combined error in the $x$- and $y$-axis is made.
}
\end{figure*}

{}

\end{document}